\documentclass[10pt,journal,compsoc]{IEEEtran}

\usepackage{color}

\ifCLASSOPTIONcompsoc
  \usepackage[nocompress]{cite}
\else
  \usepackage{cite}
\fi

\ifCLASSINFOpdf
   \usepackage[pdftex]{graphicx}
   \graphicspath{{../pdf/}{../jpeg/}{./figs/}{./figures}}
\else
\fi

\usepackage{graphicx}
\ifCLASSINFOpdf

\else

\fi

\usepackage[cmex10]{amsmath}
\usepackage{array}
\usepackage{multirow}
\usepackage{lineno}
\usepackage{url}
\usepackage{amsmath}

\ifCLASSOPTIONcompsoc
  \usepackage[nocompress]{cite}
\else
  \usepackage{cite}
\fi

\usepackage{framed,multirow}
\usepackage{amsmath}
\usepackage{amssymb}
\usepackage{latexsym}
\usepackage{url}
\usepackage{xcolor}
\usepackage{hyperref}
\usepackage{booktabs}
\usepackage{color}
\usepackage{graphicx}

\usepackage{epsfig}
\usepackage{tabularx}
\usepackage{makecell}
\usepackage{mathtools}
\usepackage[numbers,sort&compress]{natbib}

\usepackage{hyperref}

\usepackage{xr}
\usepackage{bm}

\graphicspath{{./Figs/}}

\def\0{{\bf 0}}
\def\1{{\bf 1}}

\def\etal{{\em et al.}}
\def\eg{{\em e.g.}}
\def\ie{{\em i.e.}}

\def\etal{{\em et al.\/}\,}

\definecolor{newcolor}{rgb}{.8,.349,.1}

\begin{document}
\title{Unpaired Volumetric Harmonization of Brain MRI with Conditional Latent Diffusion} 

\author{Mengqi~Wu,
Minhui~Yu,
Shuaiming~Jing, 
Pew-Thian~Yap,
Zhengwu~Zhang,
Mingxia~Liu,~\IEEEmembership{Senior Member,~IEEE}

\IEEEcompsocitemizethanks{
\IEEEcompsocthanksitem M.~Wu, M.~Yu, S.~Jing, P.-T.~Yap, and M.~Liu are with the Department of Radiology and Biomedical Research Imaging Center (BRIC), University of North Carolina at Chapel Hill, Chapel Hill, NC 27599, USA. 
M.~Wu and M.~Yu are also with the Joint Department of Biomedical Engineering, University of North Carolina at Chapel Hill and North Carolina State University, Chapel Hill, NC 27599, USA. 
Z.~Zhang is with the Department of Statistics and Operations Research, University of North Carolina at Chapel Hill, Chapel Hill, NC 27599, USA. 
\IEEEcompsocthanksitem 
Corresponding author: M.~Liu (Email: mxliu@med.unc.edu). 
\protect\\
}
} 
\if false
\markboth{IEEE Transactions on Pattern Analysis and Machine Intelligence}%
{}
\fi 

\IEEEtitleabstractindextext{
\begin{abstract}
Multi-site structural MRI is increasingly used in neuroimaging studies to diversify subject cohorts.
 However, combining MR images acquired from various sites/centers may introduce site-related non-biological variations. 
Retrospective image harmonization helps address this issue, but current methods usually perform harmonization on pre-extracted hand-crafted radiomic features, limiting downstream applicability. 
Several image-level approaches focus on 2D slices, disregarding inherent volumetric information, leading to suboptimal outcomes. 
To this end, we propose a novel 3D MRI Harmonization framework through Conditional Latent Diffusion (HCLD) by explicitly considering image style and brain anatomy. 
It comprises a generalizable \emph{3D autoencoder} that encodes and decodes MRIs through a 4D latent space,
and a \emph{conditional latent diffusion model} that learns the latent distribution and generates harmonized MRIs with anatomical information from source MRIs while conditioned on target image style. 
This enables efficient volume-level MRI harmonization through latent style translation, without requiring paired images from target and source domains during training. 
The HCLD is trained and evaluated on 4,158 T1-weighted brain MRIs from three datasets in three tasks, 
assessing its ability to remove site-related variations while retaining essential biological features. 
Qualitative and quantitative 
experiments suggest the effectiveness of HCLD over several state-of-the-arts. 
\end{abstract}

\if false
\begin{abstract}
Multi-site structural MRI is increasingly used in neuroimaging studies to enhance subject diversity. However, pooling MRI data from different sites introduces site-related non-biological variations, which can hinder the development of generalizable models. Retrospective harmonization methods aim to address this issue, but existing approaches have notable limitations. Non-learning image-level methods often focus on global intensity scaling, which may not fully remove site-related variations, while non-learning statistical methods typically harmonize only pre-extracted features, potentially limiting their applicability for downstream analysis. Deep-learning methods generally perform image-level harmonization but often operate on 2D slices, which can disregard volumetric information, potentially leading to artifacts or spatial discontinuities. Additionally, some of these methods, particularly those using GAN-based architectures, may be unstable and time-consuming to train.
To address these challenges, we propose a novel learning-based 3D MRI Harmonization framework through Conditional Latent Diffusion (HCLD). HCLD explicitly considers both image style and brain anatomy, employing a 3D autoencoder and a conditional latent diffusion model to harmonize MRI volumes at the latent level. Trained and evaluated on over 4,158 T1-weighted brain MRIs, HCLD demonstrated its effectiveness in removing site-related variations while preserving essential biological features, showing superior results and reduced computational costs compared to several state-of-the-art methods.
\end{abstract}
\fi 

}

\maketitle

\IEEEdisplaynontitleabstractindextext

\IEEEpeerreviewmaketitle

\begin{figure*}[!tp]
\setlength{\abovecaptionskip}{0pt} 
\setlength{\belowcaptionskip}{0pt}  
\setlength\abovedisplayskip{0pt}
\setlength\belowdisplayskip{0pt}
\centering
\includegraphics[width=1\textwidth]{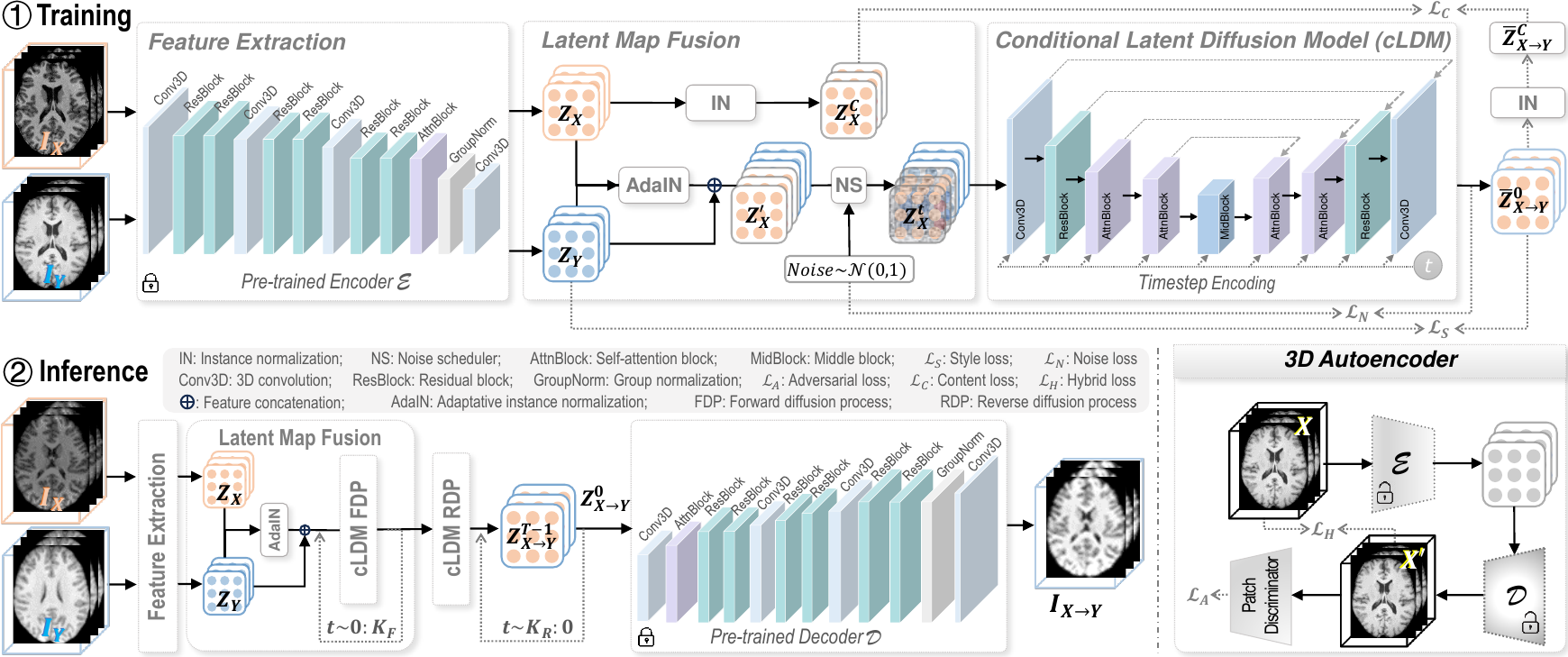}
\caption{Illustration of the proposed 
HCLD framework. 
During \emph{training}, it extracts latent feature maps from source and target MRIs using an encoder $\bm E$, fuses latent representations, and trains a conditional latent diffusion model (cLDM) to estimate the translated latent maps. 
During \emph{inference}, it applies the trained cLDM to generate the final translated latent map by iterative denoising $T_s$ steps and then utilizes a decoder $\bm D$ to reconstruct the translated MRI.
Both $\bm E$ and $\bm D$ are derived from an autoencoder pre-trained on 3,500 T1-weighted brain 
MRIs. 
}
\label{fig_pipeline}
\end{figure*}

\IEEEraisesectionheading{\section{Introduction}\label{S1}}
Neuroimaging studies increasingly utilize multi-site structural MRI to enhance subject diversity and improve the statistical power of learning-based models for purposes such as brain age-related longitudinal studies~\cite{OpenBHB,zhu2023multi_site,hawco2022longitudinal}.
However, direct pooling MRI data from various sites may introduce site-related non-biological variations that prevent models from learning generalizable features from multi-site MRIs. 
These variations, known as \emph{site/scanner effect}, can be attributed to many factors, such as differences in field strength, scanner platforms, and scanning sequences. 
Some factors, such as software and hardware updates are hard to unify across different acquisition sites~\cite{de2022magnims,wrobel2020intensity,dewey2019deepharmony}. 
Therefore, retrospective data harmonization is essential in pre-processing multi-site MRI 
to mitigate site-related variations and facilitate downstream analysis. 

Existing retrospective harmonization methods can be generally categorized as (1) non-learning and (2) learning-based methods. 
Non-learning methods can be applied directly to the image or radiomic features without training. 
Image-level non-learning methods include image-processing steps where voxel intensities of raw MRI volumes are re-scaled and standardized to a pre-defined range~\cite{min-max,statistical_norm} or to match a reference MRI scan~\cite{HM,wrobel2020intensity}. 
While these methods are fast to apply, they have limited effectiveness in removing site-related variations~\cite{li2021impact}. 
Feature-level non-learning methods, such as statistical approaches~\cite{neuroCombat,ComBat_GAM}, employ empirical Bayes models to harmonize pre-extracted MRI radiomic features (\eg, cortical thickness and gray matter volume), which may have limited applicability for downstream analysis.


Learning-based methods require proper training to capture site-related features~\cite{stamoulou2022harmonization}. 
Most of them focus on direct image-level harmonization using deep-learning approaches, such as generative adversarial networks (GANs), to translate image styles (\eg, intensity distribution, contrast, and texture) of source MRI 
to match those of a reference/target MRI. 
To preserve essential anatomical information of source MRI, some studies~\cite{CALAMITI_2021, dewey2020disentangled} employ paired T1- and T2-weighted (T1/T2-w) MRIs for model training.  
As the paired MRIs may not always be available, many recent approaches such as CycleGAN and StyleGAN utilize cycle-consistency constraints~\cite{chang_2022_cyclegan,Modanwal_2020_cyclegan,StyleGAN}
to perform style translation while retaining anatomical information without requiring paired images. 
These methods primarily harmonize 2D slices and stack them to form a final volume, leading to spatial discontinuity under different views (sagittal, coronal, and axial). 
Improving upon the single-view 2D methods, some 2.5D methods, such as ImUnity~\cite{ImUnity_2023}, combine outputs from models trained on 2D slices from different views to form the final harmonized MRI volumes. However, they still rely on slice-by-slice harmonization, which is time-consuming and neglects volumetric information.
Moreover, many existing methods require training multiple deep networks (\eg, encoder, decoder, and discriminator) simultaneously, which increases the training cost and makes the process less stable. 

To address the limitations of 2D slice-level methods and enhance the quality of harmonized MRI, this paper proposes a novel 3D MRI Harmonization framework through Conditional Latent Diffusion (HCLD) by explicitly considering image style and brain anatomy. 
As illustrated in Fig.~\ref{fig_pipeline}, the HCLD comprises two main components: (1) a generalizable \emph{3D autoencoder} that encodes brain MRIs into a 4D latent space and reconstructs MRI volumes from latent maps, and (2) a \emph{conditional latent diffusion model}~\cite{rombach2022LDM} (cLDM) that learns the latent distribution by iteratively denoising the source latent map and generates harmonized MRIs with the condition of target image style.
We utilize two-stage training for these two components.
The 3D autoencoder is first pre-trained on a large MRI dataset without requiring site labels.
In the second stage, the pre-trained autoencoder is reused with its weight frozen to encode the high-dimensional MRI data into lower-dimensional latent maps, significantly reducing the computational cost for the cLDM training.
The cLDM is trained with designated loss functions that specifically guild style translation and enforce brain anatomy preservation.
Overall, our HCLD achieves efficient volume-level MRI harmonization through latent style translation, without requiring paired training images from target and source domains.
Extensive experiments on 4,158 T1-w MRI in 3 tasks suggest the effectiveness of HCLD over several state-of-the-arts.

The major contributions of this work can be summarized as follows.
\begin{itemize}
\vspace{-4pt}
    \item We propose a new unpaired 3D harmonization method that performs volume-level style translation through a conditional latent diffusion model.  
   This method is computationally efficient and achieves higher image quality 
   compared to existing methods.
    \item We employ a two-stage training scheme that further reduces the computational cost and enhances training stability and generalizability on unseen data. 
    \item We design a latent map fusion module and specific content/style loss functions to facilitate latent style translation, improving overall image quality and brain anatomy preservation. 
    \item Our method is rigorously evaluated on three multi-site datasets with 
    T1-weighted MRIs from 4,158 subjects across three different tasks. We also experiment with various ablated model variants, different loss implementations, and different inference strategies. 
\end{itemize}

The remainder of this paper is organized as follows. 
We review the most relevant studies in Section~\ref{sec:related_work}. 
In Section~\ref{sec:method}, we introduce the details of the proposed method. 
In Section~\ref{sec:exp}, we present data involved in this work, competing methods, experimental settings, and experimental results. 
We further discuss the influence of several key components on the performance of the proposed method in Section~\ref{sec:discussion}. 
This paper is finally concluded in Section~\ref{sec:conclusion}.

\section{Related Work}
\label{sec:related_work}
\subsection{Brain MRI Harmonization} 
    %
Existing methods for brain MRI harmonization can be roughly divided into two categories: (1) non-learning methods, and (2) learning-based methods.
The non-learning methods are primarily image-processing steps applied directly to the raw MRI scans. 
These methods aim to globally normalize the voxel intensity into a pre-defined range, making MRIs from different sites more comparable. 
For example, min-max normalization~\cite{min-max} standardizes the MRI volume by simply rescaling the intensity range to $[0,1]$. 
Similarly, z-score normalization~\cite{statistical_norm} centers the intensity distribution of the MRI volume at a mean ($\mu$) of 0 and standard deviation ($\sigma$) of 1. The WiteStripe normalization~\cite{statistical_norm} goes a step further by considering brain anatomical information. 
It first calculates the $\mu$ and $\sigma$ of the normal-appearing white matter region then applies a z-score normalization to the entire volume using these values.
Besides globally standardizing the entire voxel distribution, some studies harmonize MRIs by aligning image features, such as histograms and frequency spectrum, with those of a reference MRI. 
The Histogram-Matching~\cite{HM} learns a set of standard histogram landmarks (percentiles) from the reference MRIs. It then adjusts the intensity values of input MRIs to match these landmarks using piecewise linear mapping.
Hao~\etal~\cite{guan2023domainatm} extracts the frequency spectrum of a reference MRI and replaces certain low-frequency regions of input MRIs with the corresponding regions from the reference. 
Although these non-learning methods are fast to apply, they are not effective at removing the site-related variations in the radiomic MRI feature level~\cite{li2021impact}.
Besides image-processing methods, another type of non-learning method includes statistical methods, such as ComBat~\cite{neuroCombat} and ComBat-GAM~\cite{ComBat_GAM}. 
They can be utilized to harmonize a set of hand-crafted radiomic features, such as gray matter volume and cortical thickness, extracted from pre-defined regions-of-interest (ROIs). 
These methods utilize empirical Bayes models to estimate the site-related variations, which are then removed as additive and multiplicative batch effects.
These statistical methods, while generally efficient to employ, are limited by their dependence on predefined radiomic features. 
This can restrict their applicability in downstream analyses that require additional, non-predefined MRI features.

In contrast to non-learning methods, some studies use deep-learning methods for brain MRI harmonization. 
These techniques require training on a dataset to learn parameters that can capture site-related variations.
Inspired by image style transfer in natural image analysis, recent studies have employed  
generative adversarial network (GAN) models to tackle medical data harmonization problems on the image level~\cite{chang_2022_cyclegan,Modanwal_2020_cyclegan,StyleGAN}. 
These methods engage the generator and discriminator networks in an adversarial game, where the generator creates synthetic images resembling the real dataset distribution, and the discriminator differentiates between synthetic and real images~\cite{CycleGAN2017}. 
For instance, CycleGAN introduces a cycle-consistency constraint in its loss function for unpaired image translation and content (anatomical structure) preservation~\cite{CycleGAN2017}. 
Style-encoding GAN~\cite{StyleGAN}, inspired by StarGAN-V2~\cite{StarGANv2}, further separates the content and style encoding in the latent space, allowing the site-specific style code to be learned using a separate mapping network and injected when the generator decodes the latent code back to image space.
ImUnity~\cite{ImUnity_2023} modifies the GAN structure by adding a site/scanner unlearning module to encourage the encoder to learn domain-invariant latent representations. 
These have contributed to the continual advancements of GAN-based harmonization methods.

In addition to GAN-based models, recent studies have introduced an alternative approach that employs encoder-decoder networks to disentangle anatomical and contrast information in latent space for MRI harmonization. 
For instance, CALAMITI~\cite{CALAMITI_2021} first uses T1- and T2-weighted (T1/T2-w) MRI pairs to learn global latent codes containing anatomical and contrast information, and then disentangles 
 style and content latent codes via separate encoders and decoders.  
Dewey~\etal~\cite{dewey2020disentangled} leverage T1-w and T2-w image pairs to attain a disentangled latent space, comprising high-dimensional anatomical and low-dimensional contrast components via a Randomization block. This block allows 
generating MRIs with identical anatomical structures but varying contrast. 
Zuo~\etal~\cite{zuo2022disentangling} enhance this approach without requiring 
paired MRI sequences. 
They employ 2D slices from axial and coronal views of the same MRI to provide the same contrast but different anatomical information. 

However, current image-level methods typically harmonize 2D slices and then stack them to create a final harmonized volume. 
This approach may cause artifacts and spatial discontinuities across different views (sagittal, coronal, and axial). 
Some 2.5D methods, like ImUnity~\cite{ImUnity_2023}, merge outputs from models trained on 2D slices from various perspectives but still perform slice-by-slice harmonization, overlooking inherent volumetric information of 3D MRIs.
While some GAN-based 2D methods can be adapted for 3D data, they often face challenges in training due to instability~\cite{jung2021conditional, croitoru2023diffusion}. 


\if false
{\color{red}
CycleGAN 
function primarily for one-to-one MRI harmonization, which means they need to be fine-tuned for each source-target site pair or completely retrained when applied to new data~\cite{ImUnity_2023}. 
StarGAN treats each scan as a unique domain with its own style and harmonizes the source images to a single target scan~\cite{StyleGAN}. This approach may fail to capture the global distribution of the target site and may ignore the intra-site variations. 
Moreover, certain existing studies necessitate paired MRI scans during
training, like traveling subjects or multiple MR sequences,
which are often unavailable in retrospective studies. 
Additionally, some methods involve simultaneous training of multiple encoders, decoders, or sub-networks for separated style and content encoding, resulting in increased computation and time cost for model training. 
Besides, many of these approaches suffer from limited image generation power due to mode collapse. Thus, our proposed framework aims to enhance both the efficiency of image-level harmonization and the model's ability to generalize to independent data, while also expanding its capacity for site-specific image synthesis.
}
\fi 

\subsection{Diffusion Models}
Denoising diffusion probabilistic models (DDPMs)~\cite{ho2020DDPM} have caught much attention in the deep-learning field as a better alternative to GAN models for generative tasks.
While GANs suffer from inherent problems such as unstable training processes and mode collapse~\cite{jung2021conditional,croitoru2023diffusion}, 
diffusion models have shown good performance in image generation~\cite{xia2024diffusion,pinaya2022brain,dhariwal2021diffusion}, image inpainting~\cite{kim2022diffusionclip,avrahami2022blended}, super-resolution~\cite{saharia2022image,wu2023hsr,wang2023inversesr}, and cross-modality image synthesis~\cite{zhu2023make,jiang2023coladiff}.

A DDPM is a type of diffusion probabilistic model consisting of a forward diffusion process (FDP) and a reverse diffusion process (RDP).
The FDP is implemented as a fixed Markov Chain where a pre-defined variance scheduler adds noise to an input image, gradually destroying the image information until it becomes a complete Gaussian distribution after a fixed $T$ steps.
Conversely, the RDP is a learned Markov Chain to gradually recover the image distribution by iterative denoising from the Gaussian distribution. 
Existing DDPMs are typically implemented using a time-conditioned UNet backbone~\cite{ho2020DDPM,song2020ddim,rombach2022LDM} and trained to predict noise using a re-parameterized Gaussian transition.
Song \etal~\cite{song2020ddim} propose a denoising diffusion implicit model (DDIM), which alters the RDP as a non-Markovian sampling process while keeping the original FDP in DDPM. 
This RDP becomes a deterministic mapping from the noisy latent to images, allowing a lossless inversion of the FDP with fewer sampling steps.
Rombach \etal~\cite{rombach2022LDM} further embrace the idea of two-stage training, by first training an autoencoder to compress the high-dimensional image data into a lower-dimensional latent space. 
Following this, a latent diffusion model (LDM) is trained for subsequent generative tasks. 
The autoencoder greatly reduces the computational cost~\cite{rombach2022LDM,zhu2023make} as it moves the diffusion operations into the latent space. 
Another key advantage is that it needs to be trained only once and can then be universally applied across multiple LDM models, even those designed for entirely different tasks. 
The LDM has demonstrated superior performance across a variety of tasks. 
It also offers a flexible conditioning mechanism for incorporating auxiliary information.

Diffusion models have been increasingly 
utilized in the field of medical image analysis. 
Pinaya \etal~\cite{pinaya2022brain} employ an LDM to synthesize new T1-weighted brain MRIs conditioned on the subject age. 
Wang \etal~\cite{wang2023inversesr} propose a super-resolution method for brain MRI, leveraging a pre-trained LDM.
Zhu \etal~\cite{zhu2023make} apply LDM for cross-modality brain MRI synthesis.
Durrer \etal~\cite{durrer2023diffusion} utilize a DDPM model for harmonizing 1.5T to 3T brain MRI slices.
In all these cases, diffusion models outperform their GAN counterparts in terms of the quality of generated images and demonstrate better scalability to 3D images. 
While the previous study by Durrer \etal~\cite{durrer2023diffusion} has made significant strides in proposing a harmonization method using DDPM, it primarily focuses on 2D slice-level harmonization and necessitates the use of paired MRIs (\ie, same subjects scanned at multiple sites)
Recognizing these limitations, we introduce an innovative approach for unpaired 3D brain MRI harmonization method using conditional latent diffusion. 
Our proposed model comprises a 3D autoencoder that can encode 3D MRIs into a lower-dimensional latent space irrespective of site information. 
Additionally, we employ a latent diffusion model that generates MRIs with the source site anatomical contents while conditioned on the style information of target MRIs. 

\section{Methodology}
\label{sec:method}
\subsection{Problem Formulation}
We formulate MRI harmonization as a conditional image reconstruction problem, where the model learns to construct MRI volumes in source domains/sites while conditioning the style information of a specific target domain. 
Given MRIs from a source domain {\small$X$} and a target domain {\small$Y$}, we first employ a pre-trained encoder {\small$\bm{E}$} to map MRIs from image space to a latent space via {\small{$\bm E$$:$$~\{I_X,I_Y\}$$\to$$\{Z_X, Z_Y\}$}}. 
In this latent space, the latent map 
{\small{$Z$$=$$(Z^S,Z^C)\in \mathbb{R}^{c \times w \times h \times d}$}}, encapsulates both the MRI style {\small$Z^S$} and content {\small$Z^C$} (anatomical information). 
Here, $c$ is the number of feature channels and $w$, $h$, and $d$ represent latent dimensions.
Our goal is to train a latent diffusion model that takes the source latent content map as input and the target latent map as a condition to generate a translated latent map containing the target's style and the source's content information. 
This translation can be formulated as: 
{\small{$\bm T$$:$$~\{Z_Y$$=$$(Z^S_Y,Z^C_Y),Z^C_X\}$$\to$$\{Z_{X\to Y}$$=$$(Z^S_Y,Z^C_X)\}$}}. 
Finally, we utilize a pre-trained decoder $\bm D$ to map the translated latent map to the translated MRI, which can be formulated as: {\small{$\{Z_{X\to Y}$$=$$(Z^S_Y,Z^C_X)\}$$\to$$\{I_{X\to Y}\}$}}.

\subsection{Model Training}
\label{sec:training}
As shown in the top of Fig.~\ref{fig_pipeline}, the training process of the proposed HCLD comprises three components: (1) a feature extraction module,  which extracts deep image features from MRI volumes of the source and target domains; (2) a latent map fusion module, which combines and pre-aligns the latent feature maps of the two domains; and 3) a conditional latent diffusion module (cLDM), which learns to reconstruct source feature maps conditioned on the target style. 
Notably, only the cLDM undergoes updates during the training stage.

\subsubsection{Feature Extraction}
The feature extraction module consists of an encoder $\bm E$, which is part of a pre-trained 3D autoencoder. 
Specifically, it consists of 3 sets of residual blocks and 3D convolutional downsampling blocks, designed to reduce the spatial dimension while preserving essential image features.
The encoder $\bm E$ takes the original MRI volumes, $I_X$ and $I_Y$, from the source and target domains as input and extracts deep image features, resulting in $Z_X = \bm E(I_X)$ and $Z_Y = \bm E(I_Y)$, where $Z \in \mathbb{R}^{c \times w \times h \times d}$ is a multi-channel 4D feature map.

\subsubsection{Latent Map Fusion} 
The latent map fusion module processes the encoded feature maps $Z_X$ and $Z_Y$ through two distinct branches. 
In the top branch, an instance normalization (IN) layer standardizes {\small$Z_X$} across spatial dimensions using channel-wise mean and variance, producing {\small$Z^C_X$}. This can be expressed as:
\begin{equation}
\label{eq:IN}
Z^C_{Xi} = \text{IN}(Z_{Xi})=\frac{(Z_{Xi}-\mu(Z_{Xi}))}{\sigma(Z_{Xi})},
\end{equation}
where $i$ denotes the $i$-th channel of the source latent map. 
Previous studies show that channel-wise statistics in latent feature maps can encapsulate the style of images~\cite{gatys2016style,li2016combining,li2017demystifying,garg2023neural}. 
By standardizing each feature channel to zero mean and unit variance, the IN layer removes instance-specific style from an image while retaining \emph{essential content features} in {\small$Z^C_X$}~\cite{huang2017arbitrary}. 
Using this approach, we can get a latent representation of the content information in source MRI to reduce the influence of the source MRI style.

In the bottom branch, we utilize the Adaptative Instance Normalization (AdaIN)~\cite{huang2017arbitrary} to coarsely align the channel-wise statistics (\ie, mean and standard deviation) of the source feature map with the target's. 
And the coarsely-aligned feature map can serve as an initialization for fine-grained style transfer. 
Following~\cite{huang2017arbitrary}, we utilize the AdaIN to align the source feature map with the style of the target feature map, which can be expressed as:
\begin{equation}
\label{eq:AdaIN}
\begin{split}
Z'_{Xi} &= \text{AdaIN}(Z_{Xi},Z_{Yi})\\
&=\sigma(Z_{Yi})\frac{(Z_{Xi}-\mu(Z_{Xi}))}{\sigma(Z_{Xi})}+\mu(Z_{Yi}),
\end{split}
\end{equation}
where $i$ is the channel index.
This provides a coarsely-aligned source-to-target feature map for subsequent diffusion model training.

Subsequently, the coarsely-aligned latent map $Z'_X$ undergoes a forward diffusion process (FDP). An FDP is a fixed Markov Chain where a noise scheduler gradually adds Gaussian noise $\epsilon$ to $Z'_X$ for $t \in [1, T]$, resulting in a series of noisy source latent maps $\{Z^1_X, \cdots, Z^T_X\}$, which eventually becomes a pure Gaussian distribution. During training, starting with the original coarsely-aligned source latent map $Z^0_X = Z'_X$ and a randomly chosen time-step $t \sim T$, we can sample a \emph{noisy source latent map} $Z^t_X$ from:
\begin{equation}
\label{eq:FDP}
\begin{split}
q(Z^t_X|Z^0_X)&\coloneq \mathcal{N}(\sqrt{\bar\alpha_t}Z^0_X, (1-\bar\alpha_t)\bm{I}) \\
Z^t_X &\coloneq\sqrt{\bar\alpha_t}Z^0_X+\sqrt{1-\bar\alpha_t}\epsilon, ~~ \epsilon\sim\mathcal{N}(0,\bm I),
\end{split}
\end{equation}
where
{\small{$\bar\alpha_t$$\coloneqq$$\prod^t_{i=1}\alpha_i$}}, {\small{$\alpha_t$$\coloneqq$$1-\beta_t$}}, and $\beta_t$ is a pre-defined variance scheduler.
This noisy source latent map is then concatenated with the target latent map, which serves as a style condition, to be used as the input for the conditional latent diffusion module.

\subsubsection{Conditional Latent Diffusion}

The conditional latent diffusion module (cLDM) is designed to revert the FDP process by reconstructing the source latent map from the noisy latent maps through a series of ``denoising'' operations.
Specifically, given a noisy source latent map $Z^t_X$ at a random time-step $t$, the cLDM learns a Gaussian transition parameterized by $\small p_\theta(Z^{t-1}_X|Z^t_X)$ with a learned mean and fixed variance~\cite{ho2020DDPM}:
\begin{equation}
\small
\label{eq:ddpm_rdp}
\begin{split}
    p_\theta(Z^{t-1}_X|Z^t_X)& \coloneq\mathcal{N}(\mu_\theta(Z^t_X, Z_Y, t), \sigma^2_t\bm{I}),\\
    Z^{t-1}_{X}& =\frac{1}{\sqrt{\alpha_t}}(Z^t_{X}-\frac{1-\alpha_t}{\sqrt{1-\bar\alpha_t}}\epsilon_\theta(Z^t_X,Z_Y,t)) + \sigma_t\bm{z},
\end{split}
\end{equation}
where $\sigma^2_t=\beta_t$ is the same variance scheduler used in the FDP in Eq.~\eqref{eq:FDP} and $\bm{z}\sim\mathcal{N}(0,\bm{I})$ is an independent standard Gaussian noise. $\epsilon_\theta(Z^t_X,Z_Y,t)$ represent outputs of a deep neural network optimized using a noise-level loss:
\begin{equation}
\label{eq:L_N}
\begin{split}
\mathcal{L}_N &= \lVert\epsilon -\epsilon_\theta(Z^t_X,Z_Y,t) \rVert^2_2\\
&= \lVert \epsilon -\epsilon_\theta(\sqrt{\bar\alpha_t}Z^0_X+\sqrt{1-\bar\alpha_t}\epsilon,Z_Y,t) \rVert^2_2,
\end{split}
\end{equation}
where $\epsilon$ is the true noise added during FDP in Eq.~\eqref{eq:FDP} and $\epsilon_\theta$ represents the noise estimated by the cLDM given the current time step $t$ and noisy source latent map $Z^t_X$ as input as well as the target latent map $Z_Y$ as conditioning.

According to Eq.~\eqref{eq:ddpm_rdp}, to get the final translated latent map {\small{$Z_{X\to Y}=\bar Z^0_X$}} requires sampling iteratively through a reverse diffusion process (RDP) for {\small{$t=T_S:0$}}, which makes the training process less efficient. 
As discussed in~\cite{ho2020DDPM}, deriving from Eq.~\eqref{eq:FDP}, we can directly estimate {\small{$\bar Z_{X\to Y}$}} using the noise predicted by cLDM at any given time step $t$ through 
\begin{equation}
\label{eq:estimate}
\begin{aligned}
 \bar Z_{X\to Y} &\approx Z_{X\to Y} \\ & =\bar Z^0_X 
  =\frac{1}{\sqrt{\bar\alpha_t}}(Z^t_X-\sqrt{1-\bar\alpha_t}\epsilon_\theta(Z^t_X,Z_Y,t)). 
 \end{aligned}
\end{equation}
Since this {\small{$\bar Z_{X\to Y}$}} is a close estimate of the final translated latent map, we can then employ separate style and content constraints to ensure {\small{$\bar Z_{X\to Y}$}} is closer to {\small$Z_Y$} in style and {\small$Z_X$} in content~\cite{garg2023neural,gatys2016style,li2017demystifying,huang2017arbitrary}.
The \emph{content loss} $\mathcal{L}_C$ is the mean square error (MSE) between the content feature maps of the original source MRI, {\small{$Z^C_X$}} and the estimated harmonized MRI {\small{$\bar Z_{X\to Y}$}}, which is formulated as:
\begin{equation}
\label{eq:content_loss}
    \mathcal{L}_C = \frac{1}{c\times M}\sum\nolimits^c_{i=1}\sum\nolimits^M_{j=1}{(Z^C_{X_{ij}} - IN(\bar Z_{X\to Y_{ij}})^2)},
\end{equation}
where  
{\small$M$$=$$w$$\times$$h$$\times$$d$} is the total number of features in each channel $c$. The instance normalization (IN), as introduced in Eq.~\eqref{eq:IN}, is utilized again to normalize the channel-wise statistics and eliminate the influence of style when calculating the content loss.

In this work, we define the \emph{style loss} as the MSE between feature correlations of {\small$Z_Y$} and {\small$\bar Z_{X\to Y}$}, captured by their Gram matrices {\small$G$} and {\small$A$}, respectively, formulated as:
\begin{equation}
\label{eq:style_loss}
    \mathcal{L}_{S_g} = \frac{1}{c^2}\sum\nolimits^c_{i,j=1}(G_{ij}-A_{ij})^2,
\end{equation}
where each Gram matrix (\ie, G and A) is $c\times c$ with each entry a normalized inner product between the vectorized feature maps $F$ in a channel $c$:
\begin{equation}
    G_{ij} = A_{ij}=\frac{1}{c\times M}\sum\nolimits^M_{m=1}F_{im}F_{jm}.\\
\end{equation}
These matrices represent the correlation between feature channels and intrinsically capture the style of an image~\cite{gatys2016style,li2017demystifying,gatys2017controlling}. 
Besides the Gram matrix, other style-transfer studies~\cite{huang2017arbitrary,li2017demystifying} propose using the difference in channel-wise statistics (\ie, mean and standard deviation) as the style loss. 
Additionally, some image-to-image translation studies~\cite{kim2022style,liu2023disentangling} adopt an adversarial style loss by training a discriminator to differentiate the style differences of two image domains. 
We experiment with each option and report them in Section~\ref{sec:style_loss}. 

The total loss function for training the proposed HCLD can be expressed as a combination of these losses:
\begin{equation}
\label{eq:train_total}
    \mathcal{L}=\mathcal{L}_N + \mathcal{L}_C + \alpha\mathcal{L}_{S_g}, 
\end{equation}
where $\alpha$ controls the relative contributions of the style loss and the content loss.
After training, the cLDM learns to reconstruct latent feature maps in target style and source content by predicting the time-conditioned noise.

\subsection{Model Inference}
\label{sec:inference}
Given that our priority is to preserve the anatomical structure faithfully during style translation rather than generating diverse samples, we adopt a deterministic sampling process similar to the Denoising Diffusion Implicit Model (DDIM)~\cite{song2020ddim}, which accelerates sampling speed and reduces uncertainty~\cite{pinaya2022brain,wang2023inversesr,zhu2023make}. 
Similar to the training phase, the inference of HCLD begins by extracting latent feature maps from source and target MRIs, as shown in the bottom panel of Fig.~\ref{fig_pipeline}. 
These latent maps are first fused similarly to the training stage and then fed into the trained cLDM for the forward diffusion process (FDP). 
We then add time-conditioned noise to the source latent map for $K_{F}$ steps, with $t_1 = 1$ and $t_{K_{F}}=T_{S}$ to generate a noisy source latent map, where $T_{S}$ denotes the total number of sampling steps, which is significantly smaller than the total number of training time steps. 
Unlike the noise scheduler in the training phase that adds random Gaussian noise using randomly sampled $t\sim T$, we iteratively add the learned noise for $t=1:K_{F}$ steps, which can be expressed as:
\begin{equation}
\label{eq:ddim_fdp}
    Z^{t+1}_X = \sqrt{\bar\alpha_{t+1}}\bar Z^0_{X} + \sqrt{1-\bar\alpha_{t+1}}\epsilon_\theta(Z^t_X,Z_Y,t),
\end{equation}
where {\small$\bar Z^0_{X}$} is the predicted $Z^0_{X}$ at current time step $t$, as defined in Eq.~\eqref{eq:estimate}.
The final {\small$Z^{K_{F}}_X$} is concatenated with the target latent map, which serves as the style condition, and fed into the cLDM for the reverse diffusion process (RDP).

The RDP deterministically reverses the FDP using the conditional probability learned during training. 
We obtain the final translated latent code by iterative denoising the fused latent map for {\small$K_{R}$ }steps, starting with {\small$t_{K_{R}}=T_{S}$} as the initial time step.
For each time step $t=K_R:1$, we iteratively derive the latent code of the previous time step $t-1$ through the following formulation:
\begin{equation}
\label{eq:ddim_rdp}
     Z^{t-1}_{X} = \sqrt{\bar\alpha_{t-1}}\bar Z^0_{X} + \sqrt{1-\bar\alpha_{t-1}}\epsilon_\theta(Z^t_X,Z_Y,t),
\end{equation}
This iterative process is repeated until $t=1$, resulting in the final translated latent code {\small$Z_{X\to Y}=Z^0_X$}. 
Finally, a pre-trained decoder $\bm{D}$ is used to reconstruct the translated MRI {$I_{X\to Y}=\bm{D}(Z_{X\to Y})$}. 
This process allows the model to reconstruct MRI in the style of the target domain while preserving the content of images from source domains.

An alternative inference approach is to use the DDPM inference strategy employed in many previous studies~\cite{ho2020DDPM,durrer2023diffusion,jiang2023coladiff}. 
For DDPM inference, we initiate with the original source latent map $Z^T=Z_X$ and sample sequentially for $t=T:1$ steps using Eq.~\eqref{eq:ddpm_rdp} instead of Eq.~\eqref{eq:ddim_rdp}.
In this context, $T$ represents the total number of time steps identical to the setting in the training stage.
This approach is more time-consuming than the DDIM approach because it requires iterating through all $T$ time steps.
Additionally, it may produce stochastic results due to the second term in Eq.~\eqref{eq:ddpm_rdp}. 
By default, we use DDIM in HCLD for inference in this work. 
We also compare the performance of these two inference strategies (\ie~DDIM and DDPM) in Section~\ref{sec:inf_strategy}.

\subsection{Pre-Trained Autoencoder}
Similar to the original latent diffusion model study~\cite{rombach2022LDM}, we employ an autoencoder to constitute a two-stage training process. 
In the first stage, the autoencoder is trained and validated on the OpenBHB dataset~\cite{OpenBHB} to encode a given MRI into a lower-dimensional 4D latent map and then reconstruct it back to a 3D MRI. 
A patch-based adversarial loss {\small$\mathcal{L}_{A}$} and a hybrid loss {\small$\mathcal{L}_{H}$$=$$\mathcal{L}_{R}$$+$$\mathcal{L}_{P}$$+$$\mathcal{L}_{KL}$} are used for autoencoder training to ensure accurate MRI reconstruction from latent maps~\cite{rombach2022LDM},
where {\small$\mathcal{L}_{R}$} is an $l_1$-norm based reconstruction loss, {\small$\mathcal{L}_{P}$} is a perceptual loss, and {\small$\mathcal{L}_{KL}$} is a Kullback-Leibler divergence loss.
In the second training stage, the pre-trained autoencoder networks $\bm{E}$ and $\bm{D}$ are reused with their network parameters frozen. 
Only the cLDM is updated to reconstruct the translated source latent map with the target domain style, which is computationally efficient as it operates in low-dimensional latent space.

This two-stage training approach improves the stability of the training process, as we do not update the autoencoder and the cLDM simultaneously. 
It also improves the generalizability of our model on unseen datasets. 
Since the autoencoder is trained irrespective of site specifications, it can directly encode and decode new data without fine-tuning once trained.
Therefore, our model can harmonize new data seamlessly if it serves as the source.
If the new data serves as the target domain, only the second training stage is required to fine-tune the cLDM on the new dataset.
This process is computationally efficient as it occurs in a low-dimensional latent space.

\subsection{Implementation Details} 
 
As shown in Fig.~\ref{fig_pipeline}, both $\bm{E}$ and $\bm{D}$ comprise three sets of residual blocks and upsampling/downsampling 3D convolutional layers, with $\{32,64,64\}$ filters, respectively.
It is implemented based on the AutoencoderKL module from the MONAI framework~\cite{cardoso2022monai}.
The autoencoder is trained using Adam optimizer with an initial learning rate (LR) of $10^{-4}$ and an LR rate scheduler that reduces LR on a plateau.  

The cLDM is implemented as a conditional U-Net using MONAI framework~\cite{cardoso2022monai}, which contains 
downsampling blocks, middle blocks, and upsampling blocks. %
The downsampling blocks and upsampling blocks are symmetrical, each containing one residual block and two self-attention residual blocks, with filters of $\{32,64,64\}$, respectively. 
The middle blocks contain two residual blocks and one self-attention block with 64 filters.
The cLDM is trained using Adam optimizer with similar configurations as the autoencoder's.  
Following~\cite{ho2020DDPM}, we set the total time steps $T$$=$1,000 and variance scheduler $\beta_t$ scaled linearly from $0.0015$ to $0.0195$.
We empirically set the training hyperparameter $\alpha=0.1$.
On the other hand, $T_{s}$, $K_F$, and $K_R$ are inference-phase hyperparameters that are set to 50, 30, and 10, respectively. 
We further examine the influence of these hyperparameters in Sections~\ref{sec:train_hyp} and \ref{sec:inf_hyp}.

\section{Experiment}
\label{sec:exp}
\subsection{Materials and Image Preprocessing}
\subsubsection{Datasets}
Three public datasets are utilized, including  
(1) Open Big Healthy Brains (OpenBHB)~\cite{OpenBHB}, which contains $3,984$ T1-weighted MRIs of healthy subjects from over 58 centers; (2) Strategic Research Program for Brain Science (SRPBS)~\cite{SRPBS_TS} with 99 T1-weighted MRIs from 9 healthy traveling subjects, scanned at 11 sites/settings; and (3) IXI with 559 healthy subjects scanned at 3 hospitals in London ({{\url{https://brain-development.org/ixi-dataset/}}}).
In the experiments, we follow the official training and validation data split. 
Since the OpenBHB project includes some subjects that overlap with the IXI study, we manually exclude the MRIs of these overlapping subjects from the OpenBHB dataset.
This results in a training set of $2,835$ T1-weighted MRIs and a validation set of $665$ T1-weighted MRIs, to train the 3D autoencoder and cLDM. 
We also fine-tune the cLDM component and evaluate our HCLD on SRPBS and IXI.

\subsubsection{Data Preprocessing}
All T1-weighted MRI volumes undergo minimal preprocessing using FSL ANAT pipeline~\cite{FSL}. The main preprocessing steps include standardized field-of-view (FOV) reorientation and cropping to remove unnecessary neck regions; bias field correction to correct intensity inhomogeneities; brain extraction to strip the skull; and registration to the $1mm^3$ MNI-152 template with 9 degrees of freedom.
All preprocessed MRIs are then normalized to an intensity range of $[0,1]$.
Due to hardware limitations, each MRI volume is center-cropped to have the dimension of $184\times184\times64$.

\begin{figure*}[!tp]
\setlength{\abovecaptionskip}{-4pt} 
\setlength{\belowcaptionskip}{0pt}  
\setlength\abovedisplayskip{0pt}
\setlength\belowdisplayskip{0pt}
\centering
\includegraphics[width=1\textwidth]{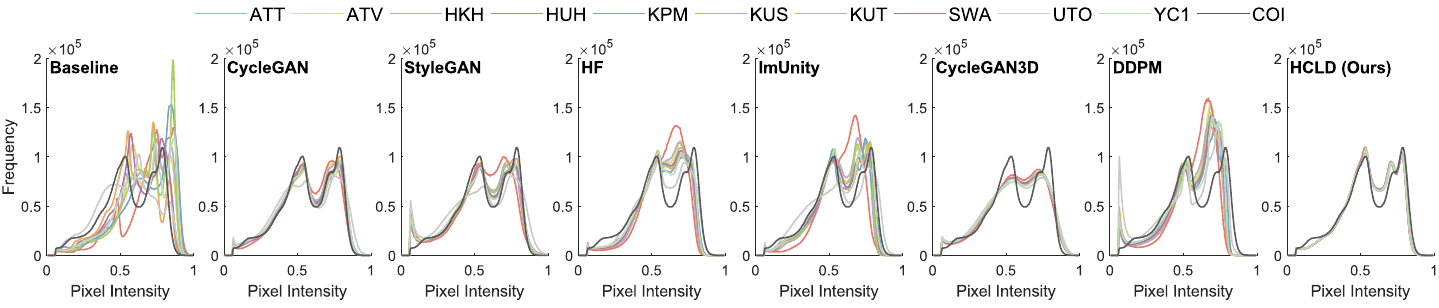}
\caption{Results of histogram comparison on 11 sites from SRPBS (with the COI site as the target domain).}
\label{fig:hist}
\end{figure*}

\begin{figure}[tp]
\setlength{\abovecaptionskip}{-4pt} 
\setlength{\belowcaptionskip}{0pt}  
\setlength\abovedisplayskip{-4pt}
\setlength\belowdisplayskip{0pt}
\centering
\includegraphics[width=0.44\textwidth]{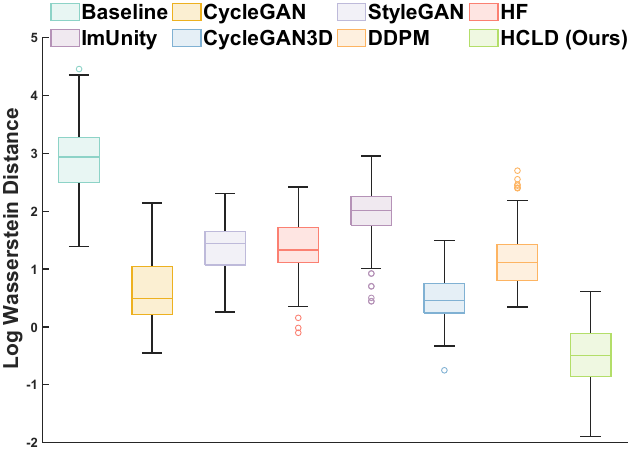}
\caption{Log Wasserstein Distance (WD) box plots showing the alignment of the sources and target histograms from the SRPBS dataset.}
\label{fig:box_visual}
\end{figure}

\begin{figure*}[!tp]
\setlength{\abovecaptionskip}{-4pt} 
\setlength{\belowcaptionskip}{0pt}  
\setlength\abovedisplayskip{-4pt}
\setlength\belowdisplayskip{0pt}
\centering
\includegraphics[width=1\textwidth]{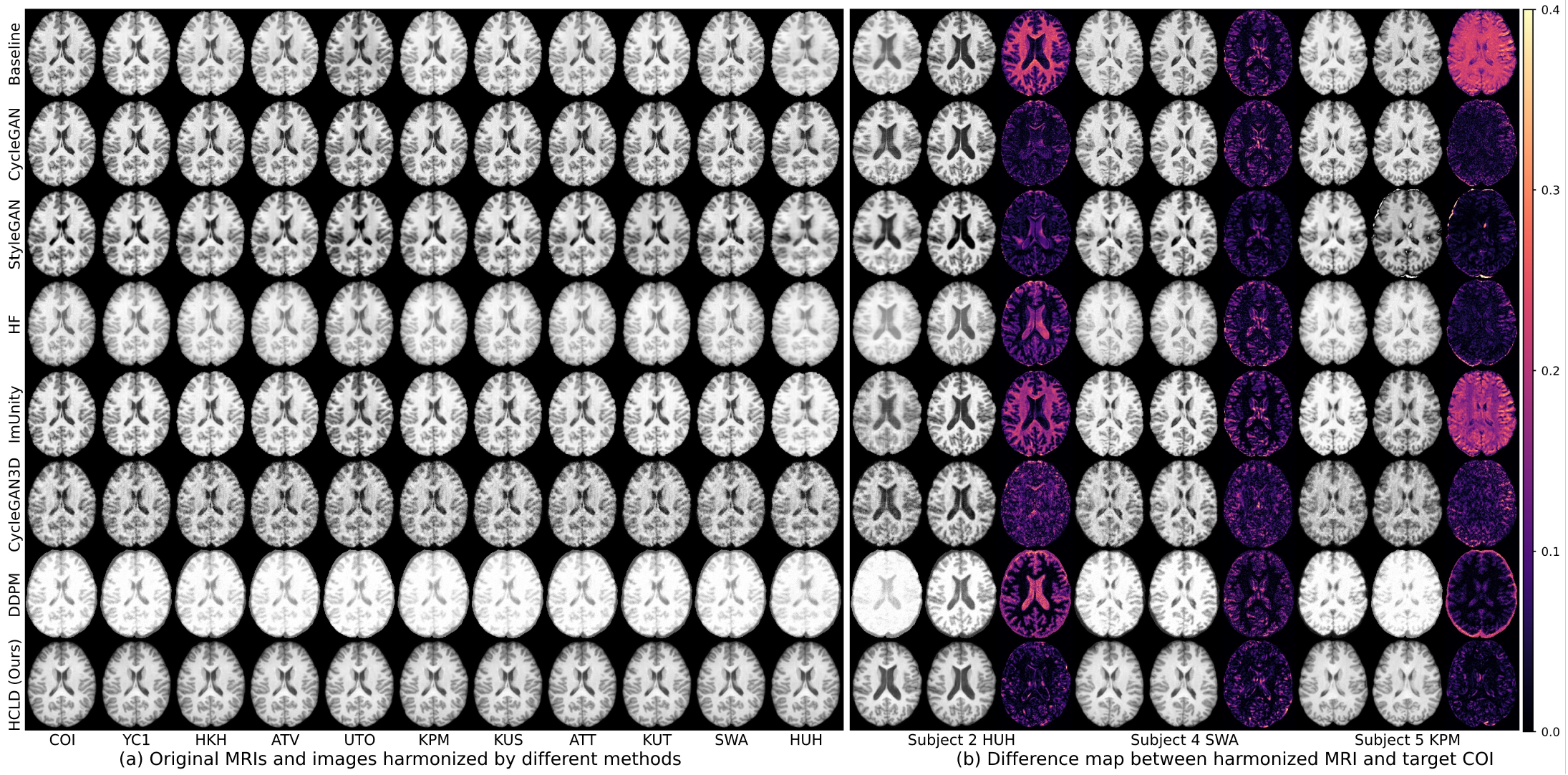}
\caption{Axial view (a) sample visualization results for SRPBS Subject 8 across 11 sites, and (b) difference map between each harmonized MRI and its ground truth for three SRPBS subjects (\ie, Subject 2 from HUH, Subject 4 from SWA, and Subject 5 from KPM).}
\label{fig:sample_site_visual}
\end{figure*}

\subsection{Experimental Settings}
\subsubsection{Competing Methods}\label{sec_competing_methods}
The proposed HCLD is compared with six methods: two 3D (\ie, DDPM~\cite{ho2020DDPM}, CycleGAN3D~\cite{CycleGAN2017}), a 2.5D (\ie, ImUnity~\cite{ImUnity_2023}), and three 2D methods (\ie, CycleGAN~\cite{chang_2022_cyclegan}, StyleGAN~\cite{StyleGAN}, and Harmonizing Flows (HF)~\cite{beizaee2023HF}). 
Details of the competing methods are specified as follows.

(1) \textbf{DDPM} method is implemented using MONAI framework~\cite{cardoso2022monai}, which comprises two downsampling blocks, a middle block, and two upsampling blocks. The downsampling and upsampling blocks are symmetrical, each containing two residual blocks and one self-attention block, with filters of $\{32,64,128\}$, respectively.
Similar to the proposed HCLD method, we concatenate source and target MRI as input to provide the model contexts of both domains. 
To maintain content information, we utilize a simple L1 pixel loss between the harmonized MRI and original source MRI.

(2) \textbf{CycleGAN3D} adopts the implementation from~\cite{ge2019unpaired}, which employs the original CycleGAN~\cite{CycleGAN2017} for 3D image harmonization. 
It comprises 2 sets of generators and 2 sets of discriminators. 
Each generator consists of three 3D convolutional layers with $\{32,64,128\}$ filters, respectively, followed by 9 residual blocks with 128 filters.
Each discriminator has five 3D convolutional layers with $\{32,64,128,256,256\}$ filters, respectively.
Both 3D methods (\ie, DDPM and CycleGAN3D) are trained using the same training and validation data as those used in the proposed HCLD method.

(3) \textbf{ImUnity}~\cite{ImUnity_2023} is specifically designed for MRI harmonization. It utilizes a VAE-GAN combined with a domain confusion module to learn domain-invariant representations and an optional biological preservation module to predict clinical-related information.
Since the data used in this work is primarily healthy control subjects, we adopt its original implementation without the optional biological preservation module. 
Following the original specification, we train 3 separate ImUnity models on 2D slices from 3 orientations (\ie,  axial, coronal, and sagittal) with the final output combined during inference, constituting a 2.5D method. 

(4) \textbf{CycleGAN}~\cite{CycleGAN2017} was initially proposed for image-to-image translation and has been applied to 2D MRI harmonization~\cite{chang_2022_cyclegan,Modanwal_2020_cyclegan}. 
We use the original implementation and train it on 2D axial slices derived from the same training and validation MRIs used in 3D methods. 
Its architecture is similar to CycleGAN3D but uses 2D convolutional layers instead of 3D ones.
After inference, the harmonized axial slices are stacked to form the harmonized MRI volumes.

(5) \textbf{StyleGAN}~\cite{StyleGAN} is a 2D MRI harmonization method implemented based on StarGAN V2~\cite{StarGANv2}. 
Utilizing the foundation of CycleGAN, it incorporates a separate mapping network and a style encoding network to learn a latent style code for each MRI and inject the learned style code into the decoder during translation. 
We adopt the default implementation and utilize the same training and inference process as described in CycleGAN.

(6) \textbf{Harmonizing Flows (HF)~\cite{beizaee2023flow}} is a recent 2D unsupervised MRI harmonization method. 
It comprises two independently trained subnetworks: an UNet-based harmonizer network, which is trained to recover MRIs from their augmented versions, and a normalizing flow network, which is trained to capture the distribution of a target domain. 
At test time, the harmonizer network is updated so that the output MRI slices match the target distribution learned by the flow network.
The original implementation trains separate models for harmonizing each source site to the target as a one-to-one translation. 
To ensure a fair comparison, we combine all source sites into a single source domain and harmonize source MRIs to a specified target domain, following the same procedure used in all competing methods.
For competing methods, we conscientiously ensure all training hyperparameters are aligned with the proposed method and that each method is trained to convergence.

\vspace{-4pt}
\begin{table*}[!t]
\setlength{\abovecaptionskip}{0pt}
\setlength{\belowcaptionskip}{-2pt}
\setlength{\abovedisplayskip}{-2pt}
\setlength\belowdisplayskip{0pt}
\centering
\scriptsize
\caption{Performance of site classification and age prediction models on harmonized MRI from OpenBHB. Values indicate mean ± standard deviation.}
\label{tab:site_age_cl}
\setlength{\tabcolsep}{5pt}
\renewcommand{\arraystretch}{0.7}
\begin{tabular*}{0.85\textwidth}
{@{\extracolsep{\fill}}l|ccc|cc}
\toprule
 \multirow{2}{*}{~~Method} & \multicolumn{3}{c|}{Site Classification}& \multicolumn{2}{c}{Age Prediction} \\
 \cmidrule{2-4} \cmidrule{4-6}
           & \multicolumn{1}{c}{BACC $\downarrow$} & \multicolumn{1}{c}{F1 $\downarrow$}
           & \multicolumn{1}{c|}{PRE $\downarrow$}& \multicolumn{1}{c}{MAE $\downarrow$} & MSE $\downarrow$\\ 
\midrule
\multicolumn{1}{l|}{Baseline}  & $0.552 \pm 0.158$& $0.650 \pm 0.122$&$0.712 \pm 0.075$ & $6.624 \pm 0.577$    &   $82.961\pm15.543$\\
\multicolumn{1}{l|}{CycleGAN~\cite{CycleGAN2017}}  &                           $0.523\pm0.054$&                         $0.642\pm0.038$ &  $0.706 \pm 0.014$  &                      $6.923\pm0.069$&   $85.625\pm2.199$\\

\multicolumn{1}{l|}{StyleGAN~\cite{StyleGAN}}    &                           $0.404\pm0.033$&                         $0.532\pm0.015$ &  $0.587\pm 0.006$&                        $7.637\pm0.060$&   $100.100\pm1.034$\\

\multicolumn{1}{l|}{HF~\cite{beizaee2023HF}}        & $0.554\pm0.067$& $0.651\pm0.060$&    $0.708\pm0.027$ &                          $6.488\pm0.083$&   $77.038\pm2.316$\\ 

\multicolumn{1}{l|}{ImUnity~\cite{ImUnity_2023}}     &                           $0.458\pm0.118$&                         $0.597\pm0.093$ &$0.667\pm 0.046$&                          $6.962\pm0.221$&   $89.349\pm8.046$\\

\multicolumn{1}{l|}{CycleGAN3D~\cite{CycleGAN2017}}  & $0.348\pm0.050$     & $0.489\pm0.029$    &$0.543\pm 0.013$&                          $6.081\pm0.027$&   $63.808\pm0.706$\\
\multicolumn{1}{l|}{DDPM~\cite{ho2020DDPM}}        & $0.451\pm0.163$& $0.574\pm0.118$&    $0.647\pm0.077$ &                          $8.174\pm0.073$&   $115.261\pm7.410$\\

\multicolumn{1}{l|}{HCLD (Ours)} & $\textbf{0.289}\pm\textbf{0.075}$     & $\textbf{0.452}\pm\textbf{0.060}$    &$\textbf{0.535}\pm \textbf{0.024}$& $\textbf{5.245}\pm\textbf{0.280}$    &   $\textbf{53.777}\pm\textbf{4.208}$\\ 
\bottomrule
\end{tabular*}%
\end{table*}

\subsubsection{Evaluation Tasks}
Three tasks are performed in the experiments, including  
(1) histogram comparison and sample visualization using the SRPBS dataset, 
(2) acquisition site and brain age classification using the OpenBHB dataset, and 
(3) voxel-level evaluation using the SRPBS and the IXI datasets.


\subsection{Result and Analysis}
 \subsubsection{Task 1: Histogram and Visual Comparison}
 \label{sec:exp1}
This experiment qualitatively assesses the results of image-level harmonization by comparing the MRI histograms from 11 SRPBS sites, both before and after the harmonization process using each harmonization method. 
We select one imaging site as our target and harmonize all MRIs from the SRPBS dataset to this target domain.
To determine a target site, we compare the intra-site variations of each site, defined as the mean peak signal-to-noise ratio (PSNR) between each pair of images within a specific site. 
Since the SRPBS dataset comprises all traveling subjects, each site contains the same subject cohort (\ie, content information).
Therefore, a site with a higher mean PSNR indicates low intra-site style variations.
In our experiment, we choose the site COI with a low intra-site variation 
as the target domain. 
We plot voxel histograms for all subjects' MRIs across 11 sites and visually compare their alignment pre- and post-harmonization using a specific method.
To quantify the harmonization effect, we also measure the difference between each source and the target (\ie, COI) histograms using Wasserstein Distance (WD)~\cite{ravano2022neuroimaging,parida2024quantitative}, which measures the amount of ``change'' required to transform one histogram into another.
To better visualize the large difference in WD results between the competing methods and the baseline, we apply the log operation to the WD results. 
In this case, a method with lower log WD denotes better histogram alignment.

Figure~\ref{fig:hist} illustrates the histogram results before harmonization (called \textbf{Baseline}) and after harmonization using seven different methods.
The Baseline highlights noticeable differences in voxel intensity distributions among each site in the raw MRI data (without harmonization) due to site-related variations.
These variations result in misaligned histogram peaks for gray matter (GM) and white matter (WM). 
Notably, our HCLD demonstrates exceptional performance in aligning histograms across all 11 sites to the histogram of the target site (depicted in black).
While CycleGAN3D and StyleGAN also align all 10 source sites, they cannot match the target intensity distribution as effectively as our HCLD.
This superior performance of HCLD may be attributed to the style alignment using AdaIN operation during latent map fusion and the diffusion model, which captures the latent data distribution of the entire target domain, instead of relying on a single reference image for style translation.
In addition, Fig.~\ref{fig:box_visual} quantitatively validates the above histogram comparison results. 
Our HCLD achieves a lower median log WD with no outliers compared to other methods, indicating better alignment of all source histograms to the target.

\begin{table*}[!tp]
\setlength{\abovecaptionskip}{0pt} 
\setlength{\belowcaptionskip}{2pt}  
\setlength\abovedisplayskip{0pt}
\setlength\belowdisplayskip{0pt}
\centering
\renewcommand{\arraystretch}{0.7}
\scriptsize
\caption{Results of volume-level evaluation on SRPBS MRIs before and after harmonization.}
\label{tab:exp3_SRPBS}
\resizebox{1\textwidth}{!}{%
\begin{tabular}{l|cccc|cccc}
\toprule
   \multirow{2}{*}{Method} & \multicolumn{4}{c|}{Intra-Site Result}& \multicolumn{4}{c}{Inter-Site Result}\\  
   \cmidrule{2-5} \cmidrule{6-9}
   &SSIM $\uparrow$ & PSNR $\uparrow$ & PCC $\uparrow$                    &WD $\downarrow$& SSIM $\uparrow$ & PSNR $\uparrow$ & PCC $\uparrow$                    &WD $\downarrow$\\ 
   \midrule
Baseline & $0.549$$\pm$$0.035$& $16.693$$\pm$$1.248$& $0.921$$\pm$$0.018$ &$0.038$$\pm$$0.032$& $0.854
$$\pm$$0.073
$& $21.754
$$\pm$$3.533
$& $0.982
$$\pm$$0.013
$&$0.041
$$\pm$$0.032
$\\
CycleGAN~\cite{CycleGAN2017}& $0.519$$\pm$$0.034$& $16.248$$\pm$$0.647$& $0.903
$$\pm$$0.015
$&$0.008
$$\pm$$0.004
$& $0.837
$$\pm$$0.073
$& $23.492
$$\pm$$2.233
$& $0.980
$$\pm$$0.014
$&$0.008
$$\pm$$0.006
$\\
StyleGAN~\cite{StyleGAN}   & $0.557
$$\pm$$0.032$& $17.091
$$\pm$$0.738
$& $0.904
$$\pm$$0.017
$&$\textbf{0.006}
$$\pm$$\textbf{0.005}
$& $0.874
$$\pm$$0.070
$& $24.280
$$\pm$$2.377
$& $0.979
$$\pm$$0.015
$&$0.009
$$\pm$$0.006
$\\
HF~\cite{beizaee2023HF}   & $0.594
$$\pm$$0.033$&$18.832
$$\pm$$0.785
$& $0.947
$$\pm$$0.009
$&$0.009
$$\pm$$0.006
$& $0.884
$$\pm$$0.063
$& $25.839
$$\pm$$2.617
$& $0.991
$$\pm$$0.007
$&$0.014
$$\pm$$0.010
$\\
ImUnity~\cite{ImUnity_2023}   & $0.567
$$\pm$$0.033
$& $16.450
$$\pm$$1.001
$& $0.924
$$\pm$$0.016
$&$0.032
$$\pm$$0.027
$& $0.874
$$\pm$$0.072
$& $22.100
$$\pm$$3.434
$& $0.983
$$\pm$$0.013
$&$0.037
$$\pm$$0.028
$\\
CycleGAN3D~\cite{CycleGAN2017} & $0.557
$$\pm$$0.032
$& $16.977
$$\pm$$0.555
$& $0.904
$$\pm$$0.013
$&$0.009
$$\pm$$0.005
$& $0.897
$$\pm$$0.070
$& $25.310
$$\pm$$2.781
$& $0.983
$$\pm$$0.014
$&$0.008
$$\pm$$0.005
$\\
DDPM~\cite{ho2020DDPM}       & $0.601$$\pm$$0.022
$& $19.061
$$\pm$$0.979
$& $0.927
$$\pm$$0.005
$&$0.014
$$\pm$$0.010
$& $0.813
$$\pm$$0.050
$& $25.596
$$\pm$$1.950
$& $0.993
$$\pm$$0.004
$&$0.013$$\pm$$0.008
$\\
HCLD~(Ours) & $\textbf{0.606}$$\pm$$\textbf{0.024
}$& $\textbf{19.367}$$\pm$$\textbf{0.674}$& $\textbf{0.951}$$\pm$$\textbf{0.008}$&$0.007
$$\pm$$0.003
$& $\textbf{0.937}$$\pm$$\textbf{0.007}$& $\textbf{29.469}$$\pm$$\textbf{0.563}$& $\textbf{0.995}$$\pm$$\textbf{0.001}$&$\textbf{0.004}
$$\pm$$\textbf{0.002}
$\\ 
\bottomrule
\end{tabular}%
}
\end{table*}

\begin{table*}[!tp]
\setlength{\abovecaptionskip}{0pt} 
\setlength{\belowcaptionskip}{2pt}  
\setlength\abovedisplayskip{0pt}
\setlength\belowdisplayskip{0pt}
\centering
\renewcommand{\arraystretch}{0.7}
\scriptsize
\caption{Results of volume-level evaluation on IXI MRIs before and after harmonization.}
\label{tab:exp3_IXI}
\resizebox{1\textwidth}{!}{%
\begin{tabular}{l|cccc|cccc}
\toprule
   \multirow{2}{*}{Method} & \multicolumn{4}{c|}{Intra-Site Result}& \multicolumn{4}{c}{Inter-Site Result}\\  
   \cmidrule{2-5} \cmidrule{6-9}
   &SSIM $\uparrow$ & PSNR $\uparrow$ & PCC $\uparrow$                    &WD $\downarrow$& SSIM $\uparrow$ & PSNR $\uparrow$ & PCC $\uparrow$                    &WD $\downarrow$\\ 
   \midrule
Baseline & $0.548$$\pm$$0.025$& $16.742$$\pm$$1.317$& $0.924$$\pm$$0.016$ &$0.034$$\pm$$0.031$& $0.549$$\pm$$0.021$& $16.561$$\pm$$1.303$& $0.928$$\pm$$0.014$ &$0.046$$\pm$$0.033$\\
CycleGAN~\cite{CycleGAN2017}& $0.570$$\pm$$0.024$& $17.348$$\pm$$1.112$& $0.940$$\pm$$0.025$ &$0.013$$\pm$$0.016$& $0.569$$\pm$$0.023$&$17.410$$\pm$$0.974$&$0.942$$\pm$$0.020$&$0.013$$\pm$$0.014$\\
StyleGAN~\cite{StyleGAN}   & $0.572$$\pm$$0.023$& $17.809$$\pm$$0.781$& $0.946$$\pm$$0.010$ &$0.007$$\pm$$0.004$& $0.574$$\pm$$0.022$& $17.868$$\pm$$0.777$& $0.947$$\pm$$0.010$ &$0.008
$$\pm$$0.004
$\\
HF~\cite{beizaee2023HF}   & $0.603
$$\pm$$0.024$& $18.614
$$\pm$$0.835
$& $0.949
$$\pm$$0.008
$&$0.008
$$\pm$$0.003
$& $0.608
$$\pm$$0.023
$& $18.532
$$\pm$$0.832
$& $0.953
$$\pm$$0.008
$&$0.008
$$\pm$$0.004
$\\
ImUnity~\cite{ImUnity_2023}    & $0.544$$\pm$$0.025$& $16.355$$\pm$$0.917$& $0.919$$\pm$$0.016$&$0.021$$\pm$$0.017$& $0.545$$\pm$$0.023$& $16.434$$\pm$$0.799$& $0.923$$\pm$$0.015$&$0.029
$$\pm$$0.018
$\\
CycleGAN3D~\cite{CycleGAN2017} & $0.602
$$\pm$$0.027
$& $18.102
$$\pm$$0.822
$& $0.952
$$\pm$$0.009
$&$\textbf{0.006}
$$\pm$$\textbf{0.003}
$& $0.603
$$\pm$$0.026
$& $18.136
$$\pm$$0.805
$& $0.952$$\pm$$0.009
$&$0.010
$$\pm$$0.005
$\\
DDPM~\cite{ho2020DDPM}       & $0.511
$$\pm$$0.024
$& $16.253
$$\pm$$0.657
$& $0.931
$$\pm$$0.011
$&$0.019
$$\pm$$0.015
$& $0.503
$$\pm$$0.023
$& $16.335
$$\pm$$0.572
$& $0.932
$$\pm$$0.010
$&$0.023
$$\pm$$0.015
$\\
HCLD~(Ours) & $\textbf{0.612}$$\pm$$\textbf{0.023}$&$\textbf{19.275}$$\pm$$\textbf{0.737}$& $\textbf{0.955}$$\pm$$\textbf{0.008}$&$0.007$$\pm$$0.006$& $\textbf{0.612}$$\pm$$\textbf{0.021}$&$\textbf{19.199}$$\pm$$\textbf{0.743}$&$\textbf{0.955}\pm$$\textbf{0.008}$&$\textbf{0.007}$$\pm$$\textbf{0.003}
$\\ 
\bottomrule
\end{tabular}%
}
\end{table*}

The qualitative analysis of sample MRIs from one subject across all 11 sites, as depicted in Fig.~\ref{fig:sample_site_visual}~(a), along with the difference map between harmonized source sites and target site COI from 3 samples in Fig.~\ref{fig:sample_site_visual}~(b), 
further validate the histogram comparison results in Figs.~\ref{fig:hist}-\ref{fig:box_visual}. 
The baseline MRI scans, before harmonization, exhibit significant variations in intensity and contrast across the different sites. 
Although most harmonization methods manage to standardize the style of the MRIs, our proposed HCLD method demonstrates superior performance by aligning the style more closely to that of the target site, COI.
Our approach also produces MRIs with significantly higher image quality than the 3D methods, such as CycleGAN3D and DDPM. 
Additionally, when compared to 2.5D and 2D methods (\ie, ImUnity, CycleGAN, and StyleGAN), the HCLD generates results with fewer artifacts. 
Among the 10 source sites, HUH presents a particularly challenging case due to its distinct deviation from the target site COI. 
Our HCLD effectively harmonizes HUH to COI, whereas most other methods fail on this site, as demonstrated by the orange line in Fig.~\ref{fig:hist} and the corresponding HUH columns in Fig.~\ref{fig:sample_site_visual}. 
More visualizations can be found in Figs.~S1-S18 of \emph{Supplemental Materials}. 
Also, Figs.~S1-S9 in \emph{Supplementary Materials} illustrate that our HCLD achieves superior harmonization outcomes in the coronal view, while some 2D methods (\eg, StyleGAN and HF) exhibit noticeable artifacts or spatial discontinuity under this view. 
This is because these methods only perform slice-by-slice harmonization in the axial view, highlighting the advantage of harmonization on the 3D volume level.


\subsubsection{\emph{Task 2: Site and Brain Age Classification}}
This experiment aims to quantitatively assess the effectiveness of the HCLD in removing site-related variations while retaining essential biological features in MRI.
We use the OpenBHB dataset with 58 acquisition sites/settings.
Similar to Task 1, we first compute the intra-site variations (\ie, mean PSNR) of each of the 58 sites in OpenBHB and select the site (Site ID: 17) with the least intra-site variation as the target site. 
We then harmonize all MRIs to the target style using HCLD and each competing method.

To evaluate the harmonization effect of each method, we extract features from harmonized MRIs utilizing a pre-trained ResNet18 network~\cite{he2016resnet} 
as a deep feature extractor, with the final fully connected layer removed and all weight frozen. 
The deep features extracted from the unharmonized raw MRIs serve as the baseline, denoted as Baseline. 
We then use the extracted deep features to train a linear logistic regression model to perform multi-class ($n=58$) classification, as well as a ridge regression model to predict brain ages. 
Following~\cite{OpenBHB}, we use 5-fold cross-validation for both regression models on the OpenBHB validation set with the regularization parameter $C\in\{0.01, 0.1, 1, 10, 100\}$.
We use balanced accuracy (BACC), F1-score (F1), and precision (PRE) to evaluate site classification performance and use mean absolute error (MAE) and mean squared error (MSE) to evaluate age prediction performance.

Results in Table~\ref{tab:site_age_cl} suggest that the raw MRIs contain significant site-related features, allowing the linear regression model to accurately distinguish between sites. 
Our HCLD effectively reduces site-related variations, making it challenging for the linear classifier to differentiate sites, as reflected by the lowest BACC, F1, and PRE values.  
Moreover, although all methods are successful in removing site-related variations, most 2D and 2.5D method negatively impacts brain age prediction performance, likely due to the anatomical discontinuity caused by stacking the slice-wise harmonization result.
While both HCLD and CycleGAN3D yield improved brain age prediction scores, the HCLD leads to more significant improvements, likely due to the content conditioning and specific content loss that aid in anatomical preservation.
On the other hand, DDPM, despite operating in 3D, results in worse age prediction scores due to its stochastic sampling process and the lack of designated style and content losses function that guides style translation and enforces anatomical preservation.

\subsubsection{\emph{Task 3: Volume-Level Evaluation}}
This experiment further calculates voxel-level image metrics pre- and post-harmonization on the SRPBS and IXI datasets.
For the IXI dataset, site IOP with the least intra-site variation is used as the target domain. 
For SRPBS, we select the same target site (\ie, COI) as in previous tasks.

\begin{figure*}[!tp]
\setlength{\abovecaptionskip}{-2pt} 
\setlength{\belowcaptionskip}{0pt}  
\setlength\abovedisplayskip{-4pt}
\setlength\belowdisplayskip{0pt}
\centering
\includegraphics[width=0.99\textwidth]{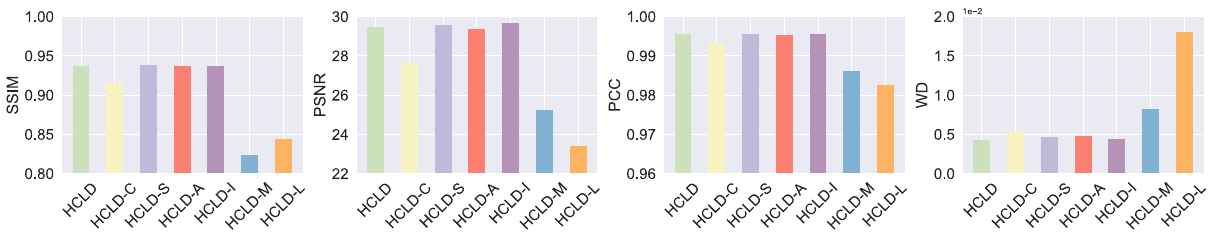}
\caption{Result of volume-level metrics of six HCLD ablation variants on MRIs from SRPBS.}
\label{fig_abl}
\end{figure*}

\begin{figure*}[!tp]
\setlength{\abovecaptionskip}{-2pt} 
\setlength{\belowcaptionskip}{0pt}  
\setlength\abovedisplayskip{-4pt}
\setlength\belowdisplayskip{0pt}
\centering
\includegraphics[width=0.99\textwidth]{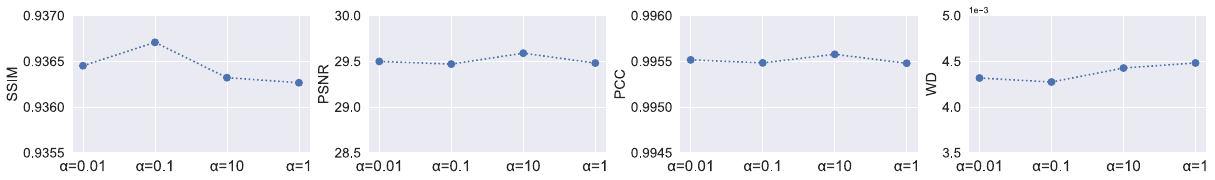}
\caption{Result of volume-level metrics of HCLD training with different $\alpha$ weights on MRIs from SRPBS.}
\label{fig_para_alpha}
\end{figure*}

We evaluate the harmonization performance using several voxel-level metrics. 
The mean structural similarity index (SSIM), intensity Pearson correlation coefficient (PCC), and peak signal-to-noise ratio (PSNR) are used to evaluate overall image quality and anatomical content integrity. 
The Wasserstein distance (WD) is used to measure style differences. 
We calculate both \emph{intra-site} and \emph{inter-site} metrics to provide a comprehensive analysis. 
Intra-site metrics are computed for every possible image pair within a single site, reflecting subject-level anatomical and image style variations within that site. 
Conversely, inter-site metrics are computed for every possible image pair between different sites, capturing both anatomical and style differences across sites.
For SRPBS which includes traveling subjects with identical anatomical information, we match subject IDs when calculating inter-site metrics. 
This allows for a direct comparison of an individual's MRI across different sites. In contrast, the IXI dataset provides a more generalized and comprehensive evaluation by considering every possible image pair.

The results in Tables~\ref{tab:exp3_SRPBS}-\ref{tab:exp3_IXI} indicate that the unharmonized data exhibit higher inter-site style variations compared to intra-site, as shown by the Baseline WD scores. Our HCLD method excels in reducing these cross-site style variations, achieving $0.004$ lower inter-site WD scores than the second-best method (\ie, CycleGAN3D) on the SRPBS dataset, and $0.001$ lower than StyleGAN and HF on the IXI dataset. 
Although some methods slightly outperform HCLD in minimizing intra-site style variations, our approach is superior in maintaining image quality and anatomical integrity, as demonstrated by the highest SSIM, PSNR, and PCC scores both inter-site and intra-site across the two datasets.

\vspace{-8pt}
\section{Discussion}
\label{sec:discussion}
\subsection{Ablation Study}
To evaluate the influence of several key components, we compared HCLD with its six simplified variants: (1) \textbf{HCLD-C} without the content loss, (2) \textbf{HCLD-S} without the style loss, and (3) \textbf{HCLD-A} without using AdaIN during latent map fusion, (4) \textbf{HCLD-I} without using IN during content loss calculation in Eq.~\eqref{eq:content_loss}, (5) \textbf{HCLD-M} that uses DDPM sampling for inference 
(instead of DDIM), 
and (6) \textbf{HCLD-L} that only decodes the result after the latent map fusion module, using the coarsely aligned latent map $Z'_X$ without the conditional latent diffusion module entirely. 
We assess all variants on SRPBS traveling subject dataset via inter-site metrics: SSIM, PSNR, PCC, and WD as used in Task 3.

Figure~\ref{fig_abl} indicates that all simplified variants lead to suboptimal harmonization results. 
Specifically, removing the content constraint (HCLD-C) leads to a notable decrease in all four metrics, suggesting a negative impact on image quality, anatomical content integrity, and style alignment.
On the other hand, removing style loss (HCLD-S) or omitting coarse latent map alignment using AdaIN (HCLD-A) mainly undermines the style translation but has little impact on the overall image quality and content integrity.
It is interesting to note that although instance normalization (IN) is used during content loss calculation, removing it (HCLD-I) primarily affects the effectiveness of style translation while leaving overall image quality and content integrity largely unaffected. 
This may be because IN normalizes the latent feature map and isolates the influence of style features during content loss calculation. 
Without IN, minimizing the content loss constrains the style change, leading to less optimal style translation, as evidenced by the higher WD score. 
Among the six HCLD variants, HCLD-L and HCLD-M experience severe performance drops across all metrics. 
This underscores the crucial role of the conditional latent diffusion module for refining the coarsely aligned latent map closer to the true target latent distribution and the substantial improvement provided by using DDIM sampling, which will be discussed in detail in Section~\ref{sec:inf_strategy}.

\subsection{Influence of Training Hyperparameter}
\label{sec:train_hyp}
We investigate the impact of the parameter $\alpha$ in Eq.~\eqref{eq:train_total} on the training process. 
This parameter regulates the balance between the style and content loss.
We conduct experiments with $\alpha\in \{0.01, 0.1, 1, 10\}$ while maintaining other parameters as constant.
As indicated in Fig.~\ref{fig_para_alpha}, the choice of $\alpha$ does not significantly impact the overall performance of the model. 
With $\alpha=0.1$, the HCLD consistently produces the highest scores across all metrics.  

\begin{figure*}[tp]
\setlength{\abovecaptionskip}{-2pt} 
\setlength{\belowcaptionskip}{0pt}  
\setlength\abovedisplayskip{-4pt}
\setlength\belowdisplayskip{0pt}
\centering
\includegraphics[width=0.99\textwidth]{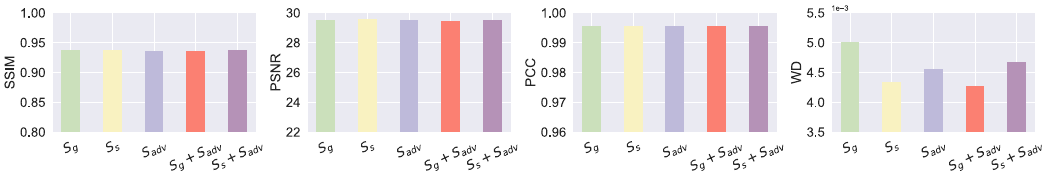}
\caption{Result of volume-level metrics of 3 style loss implementations and their combinations on MRIs from the SRPPBS dataset.} 
\label{fig_sloss}
\end{figure*}

\begin{figure*}[!tp]
\setlength{\abovecaptionskip}{-4pt} 
\setlength{\belowcaptionskip}{0pt}  
\setlength\abovedisplayskip{-4pt}
\setlength\belowdisplayskip{0pt}
\centering
\includegraphics[width=1\textwidth]{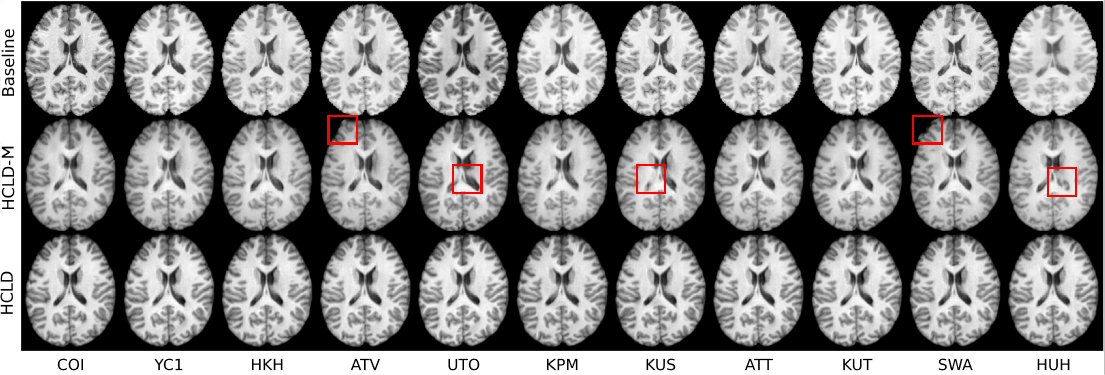}
\caption{Results of sample visualization on SRPBS achieved by the proposed HCLD (with DDIM sampling strategy) and its variant (called HCLD-M) that uses the DDPM sampling strategy during inference. Red boxes indicate areas where anatomical errors are present.} 
\label{fig:inf_stragety}
\end{figure*}

\subsection{Influence of Style Loss Implementation}
\label{sec:style_loss}
As mentioned in Section~\ref{sec:training}, there are multiple options to calculate the style loss during training. 
While the Gram matrix is used by default in HCLD, we also experiment using channel-wise statistics and adversarial learning to measure the style difference between the estimation of the translated latent map and the target latent map.
The statistical style loss is defined as:
\begin{equation}
\begin{aligned}
\small
    \label{eq:style_loss_s}
     \mathcal{L}_{S_s} = &\sum\nolimits^c_{i=1}\lVert \mu(Z_{Y_i})-\mu(\bar Z_{{X\to Y}_i})\rVert^2_2 \\ 
     &+ \sum\nolimits^c_{i=1}\lVert \sigma(Z_{Y_i})-\sigma(\bar Z_{{X\to Y}_i})\rVert^2_2,
\end{aligned}
\end{equation}
which compares the mean and standard deviation of the estimated feature map and the target feature map for each channel. 
For the adversarial style loss, we train a latent style discriminator with three 3D convolutional layers to differentiate between image domains based on latent maps. 
The style discriminator $S_D$ is trained to label real latent maps from the target domain as 1 and real latent maps from the source domain as 0. 
Simultaneously, the generator module (\ie, cLDM) is trained to fool the discriminator into classifying the translated latent maps as real target latent maps. 
A binary cross-entropy loss is used for this adversarial training, with the discriminator loss defined as:
\begin{equation}
\small
	\begin{aligned}
\mathcal{L}_{S_D} = & -\mathbb{E}_{Z_Y \sim p_{\text{data}}} [\log S_D(Z_Y)] \\
& - \mathbb{E}_{Z_X \sim p_{\text{data}}} [\log (1 - S_D(Z_X)],
	\end{aligned}
\end{equation}
and the adversarial style loss for the cLDM is  defined as:
\begin{equation}
\small
\mathcal{L}_{S_{adv}} = -\mathbb{E}_{Z_{X\to Y} \sim p_{\theta}} [\log S_D(\bar Z_{X\to Y})]. 
\end{equation}
To stabilize the training, we withhold $L_{S_{adv}}$ until after a burn-in period of 20 epochs. 
Similar to the ablation study, we calculate the voxel-level inter-site metrics on SRPBS to compare three types of style losses: (1) the statistic-based style loss $\mathcal{L}_{S_s}$, (2) the adversarial style loss $\mathcal{L}_{S_{adv}}$, and (3) the Gram matrix-based style loss  $\mathcal{L}_{S_g}$ defined in Eq.~\eqref{eq:style_loss}. 

Results in Fig.~\ref{fig_sloss} demonstrate that, while all style loss implementations uphold the same level of image quality and content integrity, the statistic-based loss $S_s$ produces the lowest WD among the individual style losses. 
And the combination of Gram-based and adversarial style loss $S_g + S_{adv}$ yields the lowest WD overall. 
One possible reason for this superior performance is that $\mathcal{L}_{S_g}$ emphasizes the similarity between low-level style features, such as intensity, captured by channel-wise correlations of the feature maps. 
On the other hand, $\mathcal{L}_{S_{adv}}$, trained on real source and target latent maps, learns to distinguish high-level stylistic features of the target domain, such as textures and patterns. 
The hybrid loss $S_g + S_{adv}$ provides comprehensive guidance for the model, leading to the optimal style alignment. 

\subsection{Influence of Inference Strategy}
\label{sec:inf_strategy}
In Section~\ref{sec:inference}, we discussed utilizing a deterministic DDIM sampling method to reduce the number of iterations required and improve anatomical preservation during inference. Here, we compare this approach with the original stochastic sampling process used in DDPM~\cite{ho2020DDPM}.
Following previous studies~\cite{ho2020DDPM,durrer2023diffusion,jiang2023coladiff} that utilize this DDPM sampling process, we sample from $t=$$~T_s$$:1$ with $T_s=T=1,000$ total steps, and denote this method as HCLD-M. 

Quantitative results from Fig.~\ref{fig_abl} demonstrate a significant decrease SSIM, PSNR, and PCC scores and increased WD, indicating reduced image quality, content preservation, and style translation.
Qualitative visualization in Fig.~\ref{fig:inf_stragety} further validates the voxel-level metrics. 
Compared to Baseline and HCLD (with DDIM sampling strategy), the HCLD-M (with DDPM sampling) shows notable anatomical errors in the cortical gray matter, ventricle, and thalamus regions, as indicated by the red boxes. 
These changes in anatomical structures during harmonization are likely due to the uncertainty introduced by the last Gaussian noise term in Eq.~\eqref{eq:ddpm_rdp}.
Therefore, we adhere to the DDIM sampling strategy for accelerated sampling and better content preservation.

\subsection{Influence of Inference Hyperparameter}
\label{sec:inf_hyp}
We further study the influence of three hyperparameters governing the DDIM sampling process, including 
(1) $T_{s}$, which controls the amount of noise added to the DDIM forward diffusion process (FDP) during the inference; 
(2) $K_{F}$ which specifies the number of iterations for the DDIM FDP; and 
(3) $K_{R}$, the number of iterations for the DDIM reverse diffusion process (RDP). 
We conduct a grid search with 10 values for each: $T_{s}\in [50, 100, 150, \cdots, 500 ]$ and $K_{F},~K_{R}\in [5, 10, 15, \cdots, 50 ]$. 
After identifying the optimal combinations, we plot the voxel-level metrics on SRPBS and visualize the trend varying one hyperparameter at a time while keeping the other two fixed. 

Line plots in Fig.~\ref{fig:ddim_hyper} illustrate the impact of varying the three hyperparameters.
The orange and blue lines denote HCLD and its variant without group normalization layers (called HCLDw/oGN), which will be discussed in Section~\ref{sec:GN}.
The two lines exhibit a similar trend in most of the plots.
Firstly, $T_{s}$ attains its optimal value at 50 steps, increasing $T_{s}$ generally leads to worse performance across all metrics.
Secondly, $K_{F}$ shows stable performance at early iterations, reaching its optimal value at 30, further increasing $K_{F}$ results in poorer outcomes across all metrics.
Lastly, $K_{R}$ has relatively less influence on the model performance.
Although the lowest WD scores are obtained at $K_{R}=25$, suggesting better style translation, we set $K_{R}=10$ as the optimal value, which leads to a higher SSIM and PSNR score, prioritizing content integrity during harmonization.

\subsection{Influence of Group Normalization}
\label{sec:GN}
A previous study~\cite{huang2017arbitrary} suggests that normalization layers, such as instance normalization (IN) and batch normalization (BN), standardize the feature maps using each sample or a batch of samples, respectively, thereby inevitably standardizing channel-wise statistics in latent feature maps. 
We have leveraged this property in Eq.~\eqref{eq:content_loss}, to reduce the influence of style information when computing content loss.
However, IN/BN layers in the final decoder of a style transfer model consistently yield worse results in their experiments because the standardization diminishes the learned channel-wise statistics, which encapsulates essential style information. 
We hypothesize that the group normalization layer (GN) used in the original cLDM and pre-trained decoder D may also be detrimental to the style translation, as they perform similar standardization on grouped feature channels.

\begin{figure}[!tp]
\setlength{\abovecaptionskip}{-4pt} 
\setlength{\belowcaptionskip}{0pt}  
\setlength\abovedisplayskip{-4pt}
\setlength\belowdisplayskip{0pt}
\centering
\includegraphics[width=0.49\textwidth]{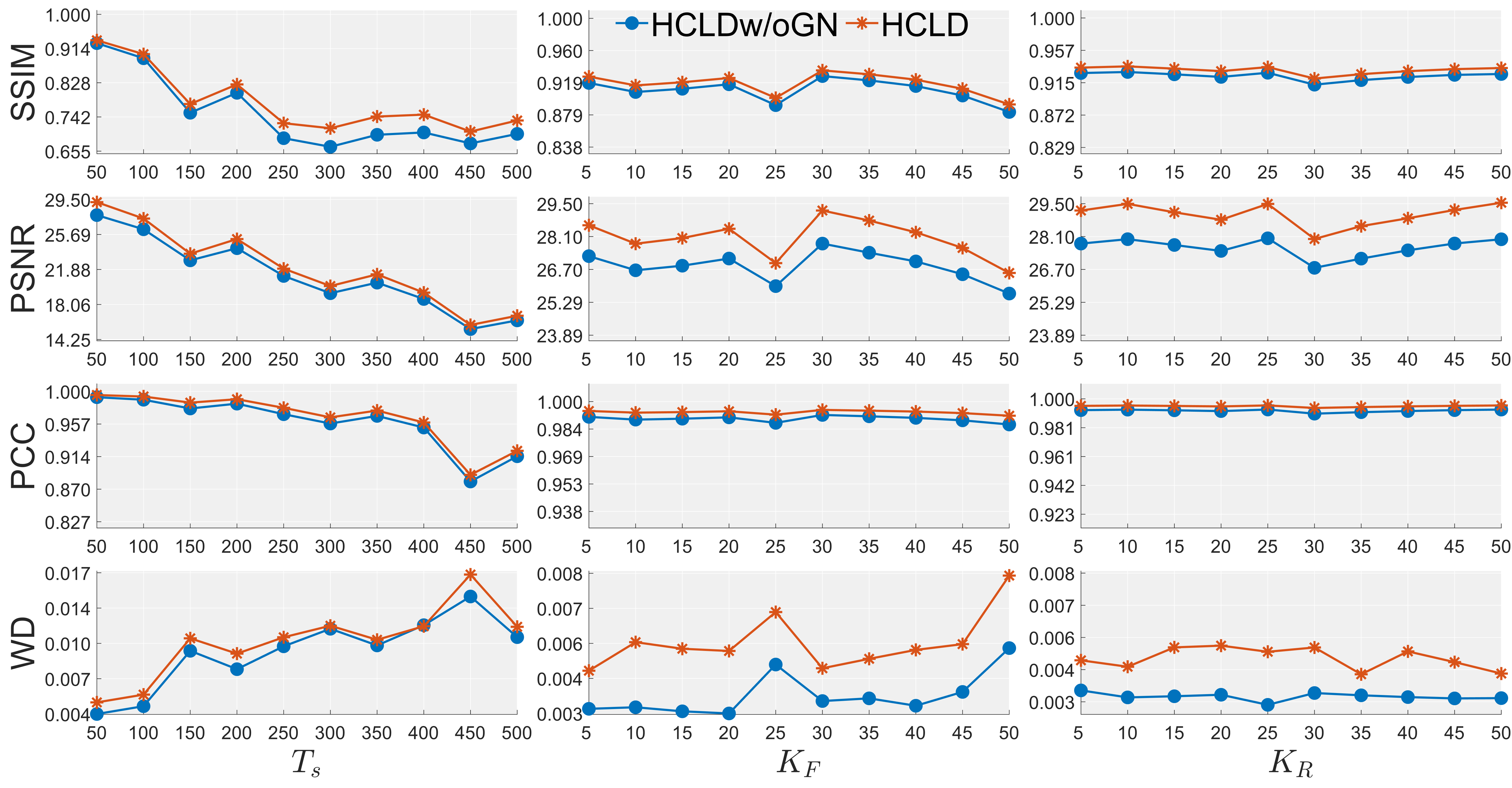}
\caption{Results of HCLD and its variant HCLDw/oGN (without group normalization layers) using different hyperparameters for DDIM inference.}
\label{fig:ddim_hyper}
\end{figure}

Line plots in Fig.~\ref{fig:ddim_hyper} substantiate our hypothesis.
The HCLD without GN layers (HCLDw/oGN), denoted by the blue line, constantly achieves a lower WD score than HCLD with GN, shown by the orange line, regardless of hyperparameter values, suggesting better style alignment overall. 
However, it is important to note that the improvement in style translation comes at the cost of overall image quality and content integrity, as the HCLD without GN shows consistently worse performance in terms of SSIM, PSNR, and PCC.
Therefore, to prioritize content integrity and image quality, we suggest keeping BN layers in the HCLD model.

\subsection{Computational Cost Comparison}
Since all the methods in this work are deep-learning based and require training, we compare their computational costs. 
We evaluate the number of trainable parameters, the total number of floating-point operations (FLOPs) in one forward pass, the total training time until convergence on SRPBS, and the inference time on SRPBS with a batch size of one.

As shown in Table~\ref{tab:cost}, our HCLD method has fewer trainable parameters than most of the competing methods and fewer FLOPs compared to other 3D methods.
It requires the least amount of training time and offers a relatively fast inference time, comparable to 2D methods (\eg, CycleGAN). 
Notably, the use of latent diffusion models and the DDIM inference strategy in HCLD significantly reduces the time costs in both the training and inference stages, compared to the DDPM method. 
These results also imply that our model is the most efficient when generalizing on a new dataset because our two-stage training strategy enables the autoencoder to be trained only once and reused on new datasets. 
Consequently, our method requires the least amount of parameters to be updated and the fewest FLOPs when fine-tuning the cLDM module on new datasets.

\begin{table}[!t]
\setlength{\abovecaptionskip}{0pt} 
\setlength{\belowcaptionskip}{2pt}  
\setlength\abovedisplayskip{0pt}
\setlength\belowdisplayskip{0pt}
\centering
\caption{Computational cost comparison across all methods. For HCLD, ``$a+b$'' denotes the number for the autoencoder and cLDM. M: Million; GMac: Giga multiply-accumulate operations; H: Hour; S: Second.}
\footnotesize
\setlength{\tabcolsep}{1.8pt}
\begin{tabular}{l|p{1.5cm}p{1.8cm}p{1.2cm}p{1.2cm}}
\toprule
     Method &  Parameters (M) &  FLOPs (GMac) &  Training Time (H) & Inference Time (S)\\ 
\midrule
     CycleGAN &  28.3 &  1,009.2 & 9.3 & 167.7\\ 
     StyleGAN &  161.3 &  4,865.3 & 10.5 & 272.4 \\ 
     HF & 5.7 & 40.5 & 48.8 & 185.3 \\ 
     ImUnity & 252.3 & 45.0 & 4.6 & 439.6\\ 
     CycleGAN3D  & 22.6 & 2,265.1 & 11.8 & \textbf{36.9}\\ 
     DDPM & 10.3 & 2,065.9 & 31.7 & 178,200.0\\ 
     HCLD (Ours) &  $3.3$$+$$\textbf{3.0}$ &  1,218.7$+$$\textbf{19.4}$ & \textbf{4.5} & 388.2 \\ 
\bottomrule
\end{tabular}
\label{tab:cost}
\end{table}

\subsection{Limitations and Future Work}
There are some limitations in the current work that can be addressed in future studies. 
\emph{On one hand}, our experiment focuses on T1-weighted MRI harmonization in healthy subjects. 
It would be more comprehensive to extend our model to include multiple MRI sequences, such as T2-weighted, T2-FLAIR, and proton-density MRIs. 
\emph{On the other hand}, beyond MRIs of healthy subjects, we can leverage the flexible conditioning mechanism enabled by the conditional latent diffusion module (cLDM) to take clinical information from patients during harmonization.
This could involve using transformers~\cite{shamshad2023transformers}
to incorporate diagnostic scores or employing spatially-adaptive normalization (SPADE)~\cite{park2019spade} blocks to utilize tissue segmentation maps, to provide additional anatomical information about the brain. 


\section{Conclusion}
\label{sec:conclusion}
This paper presents an unpaired volume-level MRI harmonization framework through conditional latent diffusion (called HCLD) with explicit content and style constraints. 
The HCLD enables efficient low-dimensional latent style translation while maintaining anatomical integrity and preserving biological features. 
Experimental results in three tasks on three datasets involving 4,158 subjects with T1-weighted MRI demonstrate the superiority of HCLD over state-of-the-art methods in aligning image style and histograms for multiple sites, eliminating site-related variations,  
and generating MR images with high quality.

\if false
\section*{Acknowledgements}
This research was supported in part by NIH grants (Nos. AG073297 and EB035160). 
\fi 

\footnotesize
\bibliographystyle{IEEEtran}
\bibliography{reference}

\if false
\vspace{-32pt}
\begin{IEEEbiography}[{\includegraphics[width=1in,height=1.25in,clip,keepaspectratio]{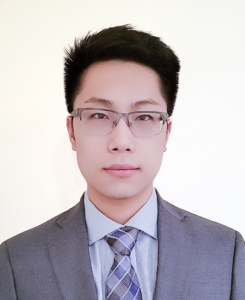}}]{Mengqi Wu}
is pursuing his Ph.D. degree in Biomedical Engineering at the University of North Carolina at Chapel Hill (2022-present). He is currently working on medical data harmonization and pattern recognition. 
\end{IEEEbiography}

\vspace{-34pt}
\begin{IEEEbiography}[{\includegraphics[width=1in,height=1.25in,clip]{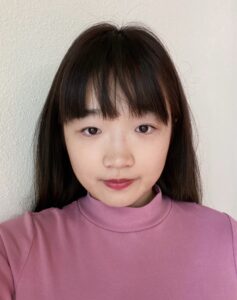}}]{Minhui Yu} is currently pursuing her Ph.D. degree in Biomedical Engineering at the University of North Carolina at Chapel Hill (2021-present). 
Her current research is mainly focused on machine learning and biomedical data analysis. 
\end{IEEEbiography}

\vspace{-34pt}
\begin{IEEEbiography}[{\includegraphics[width=1in,height=1.25in,clip,keepaspectratio]{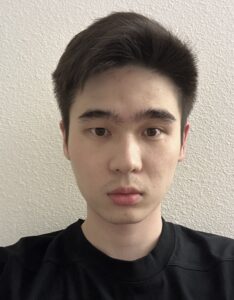}}]{Shuaiming Jing} is a graduate intern at the University of North Carolina at Chapel Hill. His current research focuses on machine learning and neuroimaging data adaptation.{\textcolor{red}{}}
\end{IEEEbiography}

\vspace{-34pt}
\begin{IEEEbiography}[{\includegraphics[width=1in,height=1.25in,clip,keepaspectratio]{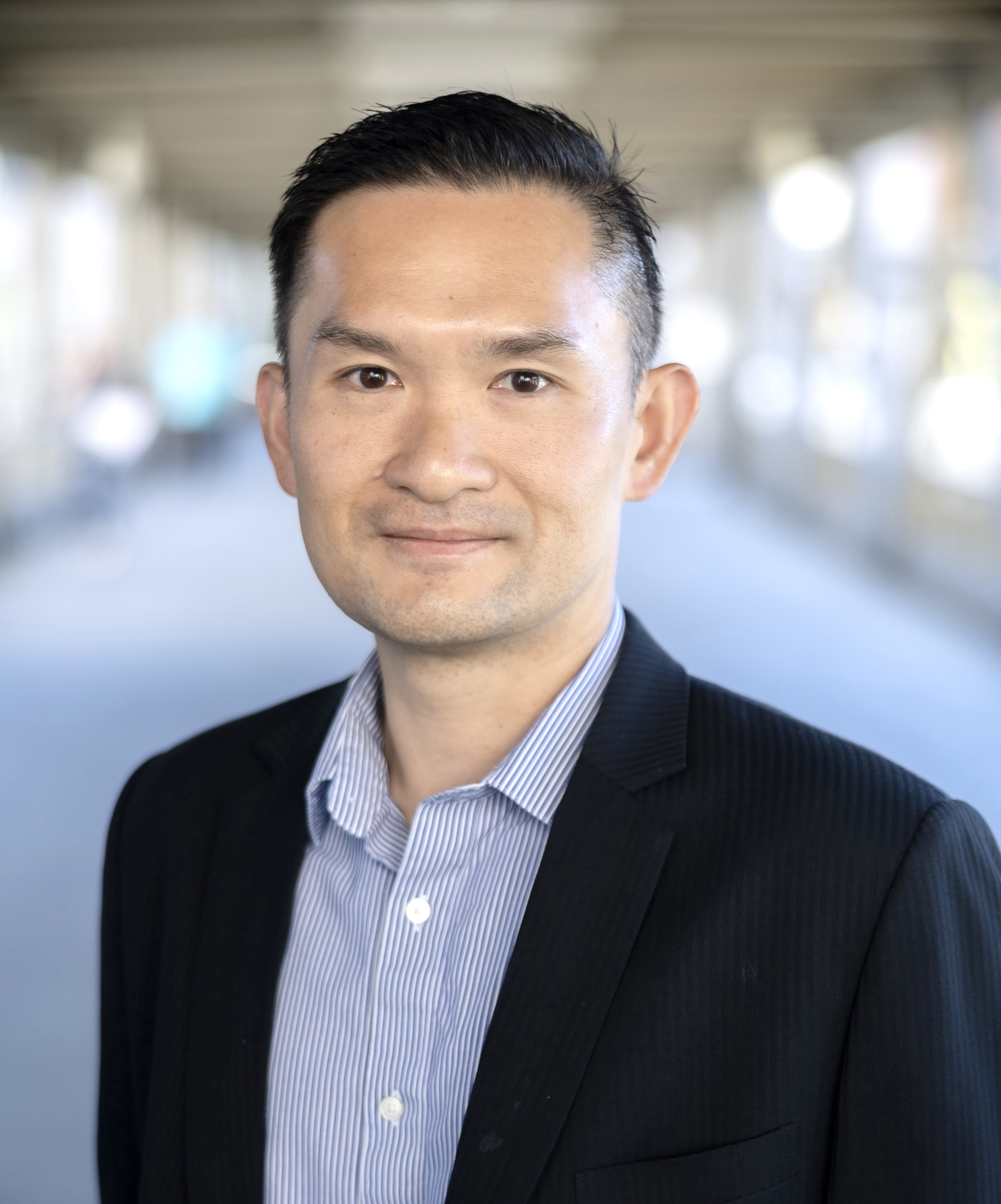}}]{Pew-Thian Yap} is a Professor of the Department of Radiology and the Director of the Image Analysis Core of the Biomedical Research Imaging Center (BRIC) at the University of North Carolina at Chapel Hill. 
His research spans image acquisition, reconstruction, quality control, harmonization, processing, and analysis. 
\end{IEEEbiography}

\vspace{-34pt}
\begin{IEEEbiography}[{\includegraphics[width=1in,height=1.25in,clip,keepaspectratio]{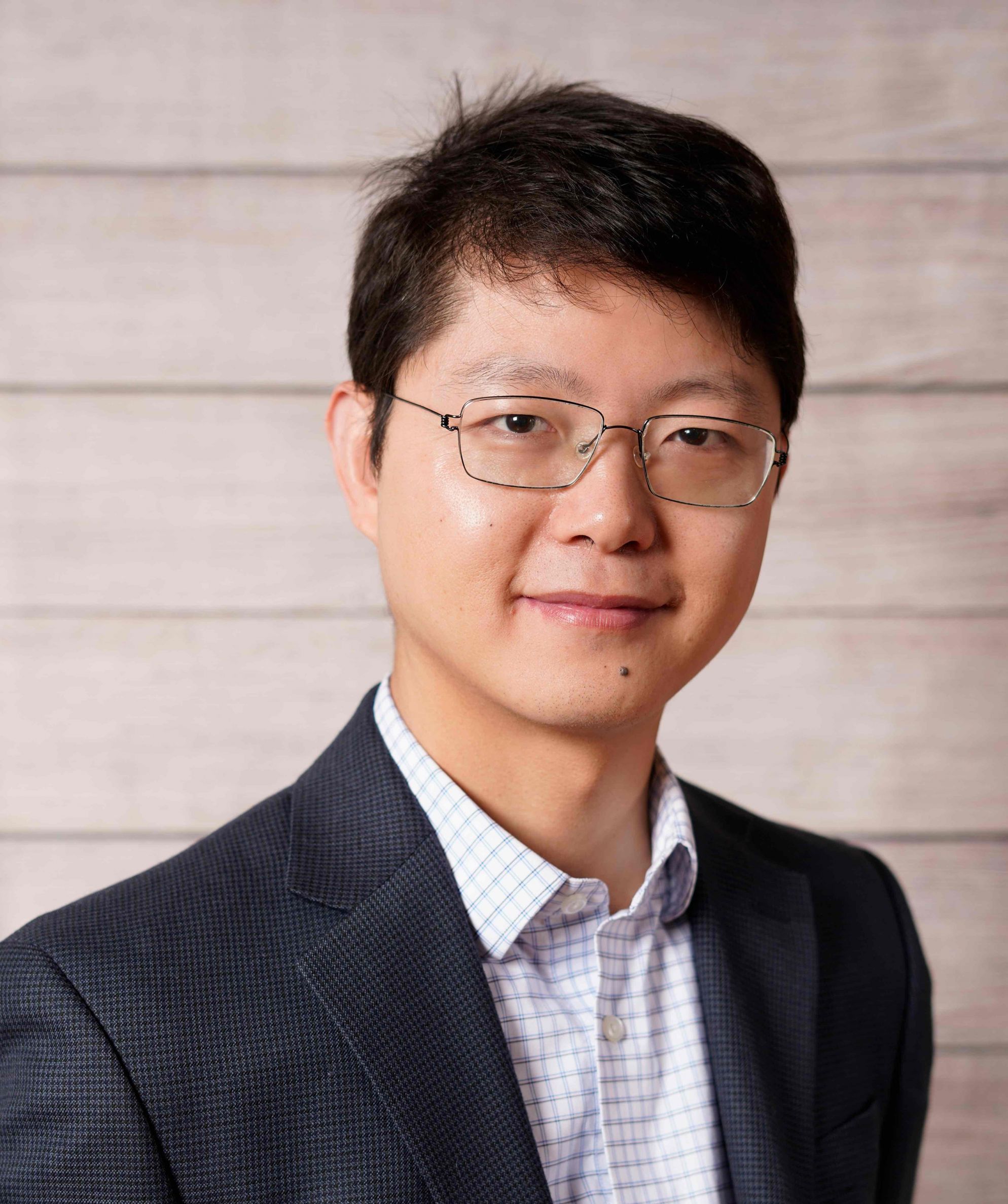}}]{Zhengwu Zhang} got his Ph.D. degree in Statistics from Florida State University in May 2015. He is currently an Assistant Professor in Statistics and Operations Research at UNC Chapel Hill. His primary research interests lie in developing effective statistical and machine learning methods for high-dimensional ``objects'' with low-dimensional underlying structures. 
\end{IEEEbiography}

\vspace{-34pt}
\begin{IEEEbiography}[{\includegraphics[width=1in,height=1.25in,clip,keepaspectratio]{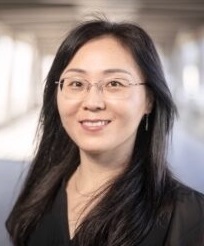}}]{Mingxia Liu}
received her Ph.D. degree in Computer Science from Nanjing University of Aeronautics and Astronautics, China, in 2015. She is a Senior Member of IEEE (SM’19). Her research interests include machine learning, pattern recognition, and medical data analysis.
\end{IEEEbiography}
\fi

\end{document}